\documentclass[manuscript]{acmart}

\usepackage{latexsym}
\usepackage[T1]{fontenc}
\usepackage[utf8]{inputenc}
\usepackage{microtype}
\usepackage{float}

\usepackage{tikz}
\usepackage{pgfplots}
\usepackage{array}
\usepackage{multirow}
\usepackage{csquotes}
\usepackage{fontawesome5}
\usetikzlibrary{backgrounds}

\usepackage[edges]{forest}
\usetikzlibrary{arrows.meta}

\usepackage{hyperref}
\usepackage{url}
\usepackage{booktabs}
\usepackage{amsfonts}
\usepackage{nicefrac}
\usepackage{microtype}
\usepackage{xcolor}

\usepackage{graphicx}
\graphicspath{ {./images/} }

\usepackage{pgfplotstable}
\usetikzlibrary{patterns}
\usepgfplotslibrary{colormaps}
\pgfplotsset{compat=1.18}
\usetikzlibrary{pgfplots.colormaps}
\usepackage{pgf-pie} 
\usepackage{xcolor,colortbl}
\usepackage{xurl}
\usepackage{siunitx}

\usepackage{dcolumn}
\newcolumntype{d}[1]{D{.}{.}{#1}}

\usepackage{makecell}
\usetikzlibrary{matrix}
\usepackage{enumitem}

\newcolumntype{L}[1]{>{\raggedright\let\newline\\\arraybackslash\hspace{0pt}}m{#1}}
\newcolumntype{C}[1]{>{\centering\let\newline\\\arraybackslash\hspace{0pt}}m{#1}}
\newcolumntype{R}[1]{>{\raggedleft\let\newline\\\arraybackslash\hspace{0pt}}m{#1}}

\AtBeginDocument{%
  \providecommand\BibTeX{{%
    \normalfont B\kern-0.5em{\scshape i\kern-0.25em b}\kern-0.8em\TeX}}}

\copyrightyear{2024}
\acmYear{2024}
\setcopyright{rightsretained}
\acmConference[FAccT '24]{The 2024 ACM Conference on Fairness, Accountability, and Transparency}{June 3--6, 2024}{Rio de Janeiro, Brazil}
\acmBooktitle{The 2024 ACM Conference on Fairness, Accountability, and Transparency (FAccT '24), June 3--6, 2024, Rio de Janeiro, Brazil}\acmDOI{10.1145/3630106.3658987}
\acmISBN{979-8-4007-0450-5/24/06}
\acmPrice{}

\begin{document}

\title{The Impact and Opportunities of Generative AI in Fact-Checking}

\author{Robert Wolfe}
\affiliation{
  \institution{University of Washington}
  \city{Seattle}
  \state{Washington}
  \country{United States}}
\email{rwolfe3@uw.edu}

\author{Tanushree Mitra}
\affiliation{
  \institution{University of Washington}
  \city{Seattle}
  \state{Washington}
  \country{United States}}
\email{rwolfe3@uw.edu}

\renewcommand{\shortauthors}{Wolfe and Mitra}

\begin{abstract}
  Generative AI appears poised to transform white collar professions, with more than 90\% of Fortune 500 companies using OpenAI's flagship GPT models, which have been characterized as ``general purpose technologies'' capable of effecting epochal changes in the economy. But how will such technologies impact organizations whose job is to verify and report factual information, and to ensure the health of the information ecosystem? To investigate this question, we conducted 30 interviews with $N$=38 participants working at 29 fact-checking organizations across six continents, asking about how they use generative AI and the opportunities and challenges they see in the technology. We found that uses of generative AI envisioned by fact-checkers differ based on organizational infrastructure, with applications for quality assurance in Editing, for trend analysis in Investigation, and for information literacy in Advocacy. We used the TOE framework to describe participant concerns ranging from the \textbf{T}echnological (lack of transparency), to the \textbf{O}rganizational (resource constraints), to the \textbf{E}nvironmental (uncertain and evolving policy). Building on the insights of our participants, we describe value tensions between fact-checking and generative AI, and propose a novel Verification dimension to the design space of generative models for information verification work. Finally, we outline an agenda for fairness, accountability, and transparency research to support the responsible use of generative AI in fact-checking. Throughout, we highlight the importance of human infrastructure and labor in producing verified information in collaboration with AI. We expect that this work will inform not only the scientific literature on fact-checking, but also contribute to understanding of organizational adaptation to a powerful but unreliable new technology.
\end{abstract}



\maketitle

\section{Introduction}

A research report issued by OpenAI in March 2023 \cite{eloundou2023gpts}, days after the release of its flagship GPT-4 model, contended that generative pretrained transformers (GPTs) are general purpose technologies, technologies with the potential to reshape not an individual profession but an entire economy. Unlike many previous general purpose technologies, the authors asserted that generative AI will impact primarily professions with a higher barrier to entry, those requiring more education and experience to carry out. Among the professions estimated by an OpenAI model as ``fully exposed'' to transformation by generative AI, defined as reducing by at least 50\% the time needed to complete the tasks of an occupation, was ``News Analysts, Reporters, and Journalists'' \cite{eloundou2023gpts}. Yet these professions have outsized epistemic effects on society \cite{ekstrom2022data}, as they remain the primary means for producing knowledge claims and for critically assessing sources and information \cite{ekstrom2021epistemologies}, thus ensuring the integrity of the online information space. If generative AI is to reshape such roles, understanding how it might do so – and where to draw the boundaries – is crucial to ensure the health of the information ecosystem. \looseness=-1 

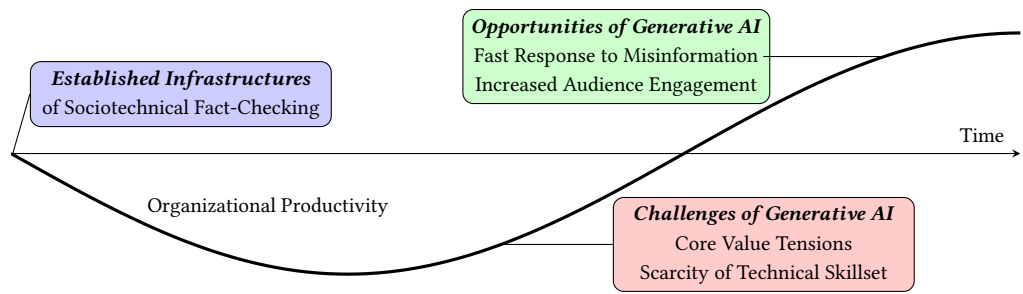
\begin{figure*}
\begin{tikzpicture}
  \begin{axis}[
    width=15cm, 
    height=8cm,
    domain=0:1.5*pi,
    samples=100,
    axis lines=middle,
    restrict y to domain=-1:1,
    ymin=-1,
    ymax=1,
    axis y line=none,
    xtick = \empty,
    xmin=0,
    xmax=1.5*pi,
  ]
    \addplot[black, very thick] {-sin(deg(x)) * .5} node[above right, pos=0.8] {};
  \end{axis}

    \node[right, draw, fill=blue!20, minimum width=4cm, align=center, rounded corners] at (.25, 4) {\small \textbf{\textit{Established Infrastructures}} \\ \small of Sociotechnical Fact-Checking};
    \node[right, draw, fill=green!20, minimum width=4cm, align=center, rounded corners] at (6, 4.5) {\small \textbf{\textit{Opportunities of Generative AI}} \\ \small Fast Response to Misinformation \\ \small Increased Audience Engagement};
    \node[right, draw, fill=red!20, minimum width=4cm, align=center, rounded corners] at (8, 2) {\small \textbf{\textit{Challenges of Generative AI}} \\ \small Core Value Tensions \\ \small Scarcity of Technical Skillset };
    \node[right] at (1.7,2.5) {\small Organizational Productivity};
    \node[right] at (12.5,3.45) {\small Time};

  \draw[-] (6.5,2) -- (8,2);
  \draw[-] (10,4.5) -- (11.62,4.5);
  \draw[-] (0.02,3.2) -- (.25,4);

\end{tikzpicture}

\caption{\footnotesize Incorporating a society-altering technology like generative AI into sociotechnical fact-checking work requires ``intangible'' investments \cite{brynjolfsson2021productivity} (new processes and skills) to realize its potential without deprioritizing the values of fact-checking or displacing the role of human experts. \looseness=-1}
\label{fig:gen_purpose_tech}

\end{figure*}

In this work, we study the impact of generative AI in fact-checking, a profession that specializes in determining the reliability of information disseminated through traditional and social media, undertaken at publishing houses and independent organizations around the world. Fact-checking is a complex sociotechnical process, involving human judgment exercised in conjunction with AI-based tools to observe misinforming claims and narratives as they spread \cite{guo2022survey}. While most fact-checking organizations necessarily embrace technological tools, they are skeptical of technologies that promise to automate large parts of the fact-checking process, and deprioritize or displace human expertise \cite{juneja2022human}. Understanding perspectives of key stakeholders at fact-checking organizations is thus important to facilitate adoption of a technology that could help respond efficiently to misinformation, while prioritizing the role of human expertise. We address two research questions: \looseness=-1

\begin{enumerate}[left=1mm]
    \item \textbf{RQ1: Opportunities of Generative AI in Fact-Checking:} What opportunities do fact-checking organizations see in generative AI? How are organizations presently using generative AI, and how do they envision using it?
    \item \textbf{RQ2: Challenges and Limitations of Generative AI in Fact-Checking:} What challenges do fact-checkers see in using generative AI to support their work? What prevents them from further incorporating generative models?
\end{enumerate}

\noindent To address these questions, we interviewed $N$=38 participants at 29 fact-checking organizations in a range of roles, from investigation, to management, to engineering. We captured diverse, global perspectives from participants located across 19 countries and six continents. Interviews provided detailed accounts of where fact-checkers envisioned using generative AI, and concrete examples of applications in use or in development. Participants also shared barriers to adopting generative AI, ranging from technical limitations to value misalignments. Figure \ref{fig:gen_purpose_tech} draws on organizational research \cite{brynjolfsson2021productivity} to illustrate the investment to overcome these challenges and realize the benefits envisioned by participants. We make four contributions: \looseness=-1

\begin{itemize}[left=1mm]
    \item \textbf{Enumerating Opportunities and Limitations of Generative AI in Fact-Checking:} We describe the opportunities for generative AI in five fact-checking infrastructures (Editing, Investigation, Audience Management, Technology, and Advocacy), and adopt the Technology-Organization-Environment framework \cite{prasad2023towards} to describe challenges.
    \item \textbf{Designing for Verification:} We propose a novel dimension in the design space for generative models that centers Verification, or ensuring the veracity of content. We describe this dimension with a 2x2 matrix, with the Producer of content on the X axis, and its Verifier on the Y axis, and discuss its use beyond fact-checking in high stakes domains. \looseness=-1
    \item \textbf{Mapping Value Tensions:} Using the principles of the International Fact Checking Network (IFCN) \cite{ifcn} as a basis for the sociotechnical values of fact-checking, we describe value tensions between fact-checking, which centers transparency and reliability, and generative AI, a technology exhibiting unpredictable and often unreliable behavior.
    \item \textbf{Defining a Research Agenda:} We propose nine directions for fairness, accountability, and transparency researchers to develop technologies, designs, and approaches supporting responsible use of generative AI in fact-checking.
\end{itemize}

\section{Related Work}

\noindent \textbf{Sociotechnical Infrastructures of Fact-Checking}. ``Fact-checking'' refers to the investigation of potentially misinforming claims and narratives that may adversely impact individuals and society \cite{graves2017anatomy, graves2019fact}.
Fact-checking is primarily a ``socio-technical'' task \cite{yu2023antecedents,radiya2023sociotechnical}, wherein technology is useful and meaningful only in the context of its relationship to the humans who interact with it \cite{tang2015restructuring, zajkac2023clinician}. While fact-checkers necessarily employ data-driven technologies \cite{guo2022survey,dierickx2023automated}, and envision further uses of technologies to, for example, minimize the amount of harmful content to which they are exposed \cite{juneja2022human}, human judgment is also crucial to the fact-checking process \cite{graves2018factsheet}, and fact-checkers are skeptical of technologies that promise to fully automate parts of fact-checking work \cite{juneja2022human}. Prior work has sought to clarify the communities \cite{brookes2023communities} and sociotechnical infrastructure undergirding the processes of fact-checking. \citet{juneja2022human} describes fact-checking organizations as composed of ``human and algorithmic infrastructures'' fulfilling distinct roles in fact-checking, such as editing and investigation. We draw on these roles when unpacking the opportunities presented by generative AI in fact-checking. \looseness=-1

\noindent \textbf{Organizational Change}. \citet{eloundou2023gpts} contend that generative AI is a ``general purpose technology'' that will reshape society and the economy, with impacts greater for professions requiring high levels of education. \citet{brynjolfsson2021productivity} describes general purpose technologies as necessitating intangible ``complementary investments'' to realize their potential, such as ``co-invention of new processes, products, business models and human capital,'' suggesting the sociotechnical nature of  technology adoption. \citet{prasad2023towards} describe generative AI's adoption using the Technology - Organization - Environment (TOE) framework, noting the impact of factors like regulation and organization size. In a now foundational text, \citet{fichman1999illusory} note that the widespread acquisition of a technology by organizations may not result in its widespread deployment, especially where ``knowledge barriers'' mitigate effective use. We borrow concepts from the TOE framework to tease apart the challenges faced by fact-checkers in adoption of this new technology. \looseness=-1

\noindent \textbf{Generative AI.} Generative AI technologies such as ChatGPT \cite{openai2022chatgpt} and its predecessors \cite{radford2018improving,radford2019language,brown2020language} can ingest human input in natural language \cite{ouyang2022training} and, depending on their architecture and training objective, produce palatable text \cite{touvron2023llama,touvron2023llama2,jiang2023mistral}, images \cite{ramesh2021zero,ramesh2022hierarchical,rombach2022high}, video \cite{wu2023tune,singer2022make}, and source code \cite{chen2021evaluating,xu2022systematic}. Such technologies both pose difficulties for fact-checkers, who must contend with higher quality misinformation produced more easily \cite{zagni2023generative,kapoor2023prepare,jain2023ai}, but also opportunities for novel technologies in their work \cite{ritala2023transforming,das2023state}. Recent work highlights difficulties with generative AI for journalism and fact-checking, including low audience trust in AI-generated content \cite{longoni2022news} and biases in the dissemination of AI-assisted fact-checks \cite{neumann2023does}. Researchers in HCI have mapped the design space \cite{card1990design} of generative AI \cite{morris2023design}, describing interactions possible with users and ways to use it in domains like scientific research \cite{morris2023scientists} and creative writing \cite{Chakrabarty2023CreativitySI}. Building on participatory design \cite{10.1145/1900441.1900448,participatorydesign}, recent work develops ``participatory AI,'' wherein human subjects envision new AI-driven designs with researchers \cite{10.1145/3617694.3623261,birhane2022power,10.1145/3593013.3594134}. For fact-checking, \citet{das2023state} conduct a review of human-centered NLP and develop a confusion matrix for calibrating trust in human-AI collaborations. We extend this line of work by proposing a socio-technical verification dimension.
\looseness=-1

\noindent \textbf{Values in AI and  Fact-Checking}. Scholars have found that AI and machine learning research is not ``value neutral'' but prioritizes values like performance and generalization, while neglecting considerations like ``negative potential'' \cite{birhane2022values}. \citet{bender2021dangers} contend that training models on poorly specified textual data poses numerous ethical risks for downstream use. Fact-checkers adhere to rigorous ethical codes \cite{doi:10.1080/17512781003642972}, such as those set forth by coalitions like the 172-member (as of this writing) International Fact Checking Network (IFCN) \cite{ifcn}, which publishes a code of principles to which its signatories commit \cite{ifcnsignatories}. We use the published principles of the IFCN, an organization formed to promote common standards in fact-checking \cite{juneja2022human}, to explore value tensions between generative AI and fact-checking. \looseness=-1

\begin{table*}[t!]
    \footnotesize
    \begin{tabular}{|L{7cm}|L{7cm}|}
    \toprule
    \multicolumn{1}{|c|}{Continents (6) | Countries (19)} & \multicolumn{1}{|c|}{Fact-checking Organizations (29)} \\
    \midrule
    \centering
   \includegraphics[height=28mm]{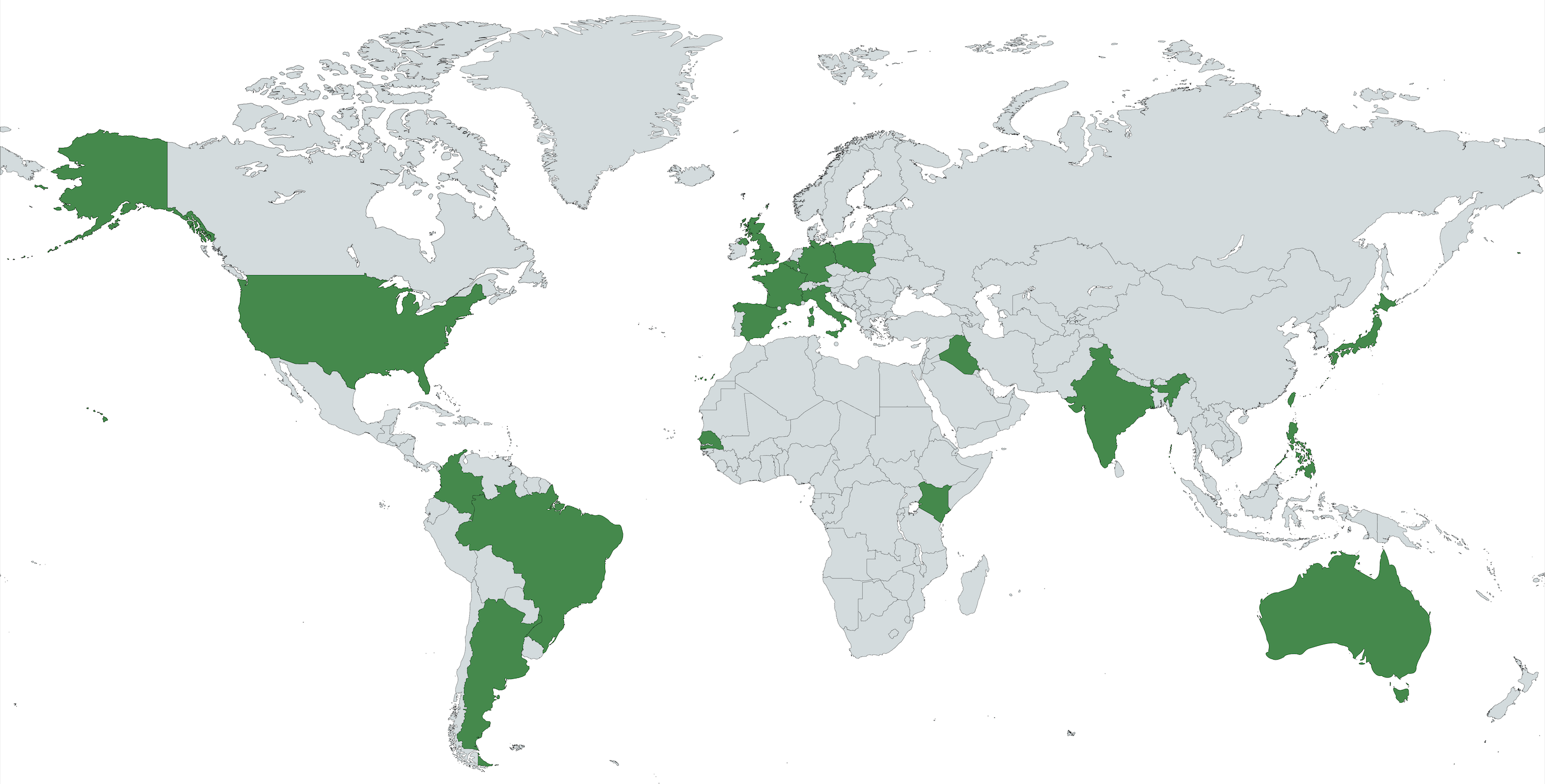} & 
    Australian Associated Press \cite{aap}, Agence France-Presse \cite{afp}, Africa Check \cite{africacheck}, Aos Fatos \cite{aosfatos}, Chequeado \cite{chequeado}, Code for Africa \cite{codeforafrica}, ColombiaCheck \cite{colombiacheck}, Der Spiegel \cite{derspiegel}, Factly \cite{factly}, India Today \cite{indiatoday}, Infoveritas \cite{infoveritas}, Lead Stories \cite{leadstories}, Litmus \cite{litmus}, logically.ai \cite{logically}, Maldita.es \cite{maldita}, Meedan \cite{meedan}, MindaNews \cite{mindanews}, Newtral \cite{newtral}, Pagella Politica \cite{pagella}, PolitiFact \cite{politifact}, Pravda \cite{pravda}, Rappler \cite{rappler}, RMIT FactLab CrossCheck \cite{rmit}, Science Feedback \cite{scifeedback}, Taiwan FactCheck Center \cite{tfc}, Tech4Peace \cite{t4p}, The Quint \cite{quint}, Thomson Reuters \cite{reuters}, Univision El Detector \cite{eldetector} \\
    \bottomrule
    \end{tabular}
    \caption{\footnotesize Participants were recruited from 6 continents, 19 countries, and 29 fact-checking organizations.}
    \label{tab:orgs}
\end{table*}

\section{Methods}

\subsection{Participant Recruitment}

We conducted an interview study with $N$=38 employees of fact-checking organizations or teams in publication houses, with experience in their current role ranging from 1 year to 18 years. As shown in Table \ref{tab:orgs}, we recruited from a total of 29 fact-checking organizations using purposive sampling and snowball sampling \cite{etikan2016comparison, naderifar2017snowball}, first reaching out to potential participants by sending an email advertising the study to the listserv of the IFCN. This resulted in three interviews with six participants working at three organizations. The Community Manager of the IFCN then provided us with contact information for six potential participants for our study, to whom we reached out. This resulted in three interviews with three participants at three organizations. We next utilized a list of 23 fact-checkers known to one of the authors, who has maintained a long-term relationship with the global fact-checking community. This resulted in five interviews with six participants at five organizations. Finally, we sent cold emails to 60 IFCN signatory organizations, explaining our interest in an interview and how we found their contact information. We recruited in this way not only to increase the number of participants in the study, but also to increase the study’s global reach, as we emailed primarily organizations in developing countries and the global south. This strategy resulted in 14 interviews with 18 participants at 14 organizations across five continents. We employed snowball sampling when participants offered to connect us with a participant well-suited to the study, and reached out via email. This resulted in five interviews with five individuals working at five organizations. \looseness=-1

\subsection{Interview Protocol}

We created a semi-structured interview protocol that posed general questions regarding the use and impact of generative AI in fact-checking. We began interviews by asking participants to tell us about their background, including their position, experience in fact-checking, and familiarity with generative AI. We asked about their company’s background, including the size and technical experience of the fact-checking team, and how long the company had been performing fact-checking work. We then explicitly posed the primary research questions of our study, asking participants to characterize 1) how they used generative AI in their work; 2) opportunities for using generative AI in fact-checking; 3) challenges and limitations of using generative AI in fact-checking; and 4) how researchers could design generative technologies that better support fact-checkers. We asked participants to clarify, discuss, and expand upon responses to better understand their perspectives. We also asked follow up questions where appropriate about several specific topics, including the use of corporate vs. open source AI; modalities (text, image, etc.) of misinformation they use generative AI to handle; use of generative AI to handle narratives; guidelines for using generative AI in their organization; and impacts the participant witnessed generative AI having. We submitted the interview protocol as part of the supporting materials to our University’s Institutional Review Board. \looseness=-1

\subsection{Interview Process}

We conducted 30 interviews between October 2023 and January 2024. Multiple participants attended seven interviews, with one participant typically a manager, and the other(s) involved in technology or investigation. In one case, we interviewed two managers who passed follow-up questions to the engineering team, forwarding their responses to us by email. Interviews lasted 30 to 90 minutes, averaging approximately 45 minutes. Interviews were conducted solely in English. We accommodated the request of one participant to send answers by email because they preferred writing over speaking in English. The participant then also met with the first author for 20 minutes. Participants who used generative AI sometimes shared their screen and displayed the interfaces used with these technologies. One participant shared a Jupyter notebook showing their use of AI in a data science pipeline. Other participants linked us to Github pages or company technical reports. We did not offer to compensate participants to prevent feelings of coercion.

\subsection{Data Analysis}

After transcribing the interviews, we deductively coded them according to four categories tied closely to the research questions: Present Use of Generative AI in Fact-Checking; Opportunities to Use Generative AI in Fact-Checking; Challenges and Limitations to Using Generative AI in Fact-Checking; and Ways Computational Research in Fairness and Transparency Can Support Fact-Checking. We then conducted inductive coding within each deductive category.

The authors first coded four interview transcripts, after which the first author created a codebook that included inductively derived codes organized within the deductive categories. The codebook consisted of the names of codes, explanations of the codes, and the associated participant quotes. The first author shared the codebook, and the last author offered feedback and suggestions, after which the authors met to discuss the codes and revised or removed codes on which they could not reach agreement. The authors then coded four additional transcripts at a time, updating the codebook after each round with more precise definitions and additional context from participant quotes. Next, the authors followed a thematic analysis process \cite{Clarke2017ThematicA,Braun2022EverythingCW} to generate themes that described the findings and addressed the research questions. Specifically, the authors reviewed the codes and their associated participant quotes, drafted memos describing proposed themes, met to discuss the proposed themes, and converged on a set of final themes on which agreement could be reached. These themes form the basis of the Findings section. 

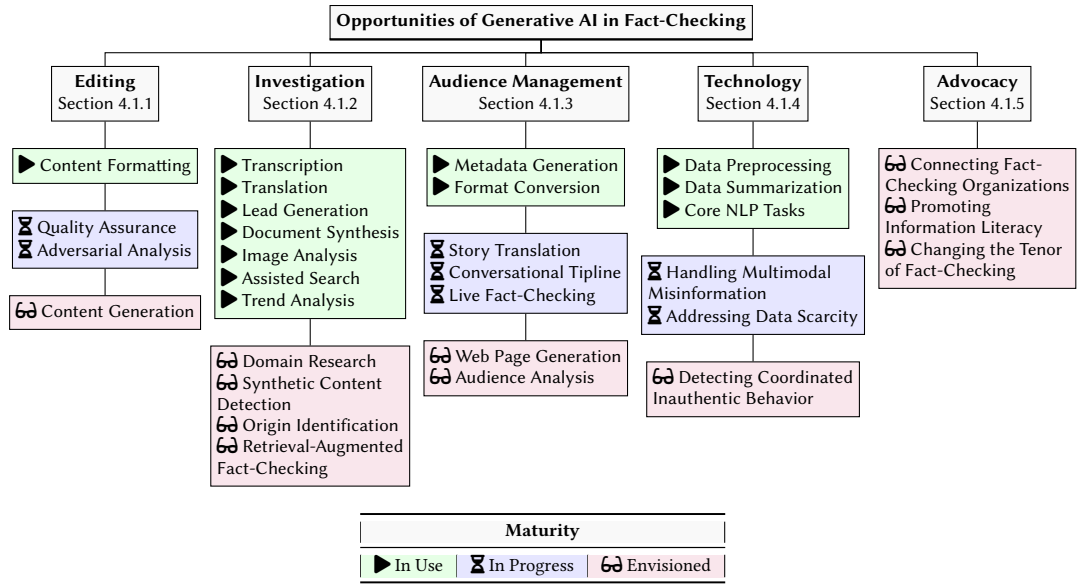
\begin{figure*}
\footnotesize
\centering
\begin{forest}
  for tree={
    align=left,
    font=\sffamily,
    l sep'=10pt,
    s sep = 5pt,
    fork sep'=5pt,
    draw,
  },
  forked edges,
  baseline,
  where level=0{
    tikz={\draw [thick] (.children first) -- (.children last);},
  }{},
  [\textbf{Opportunities of Generative AI in Fact-Checking}, fill=gray!5
    [\textbf{Editing} \\ Section \ref{sec:ed_opportunities}, fill=gray!5, align=center
      [\faPlay \ Content Formatting, fill=green!10
      [\faHourglassHalf \ Quality Assurance \\
      \faHourglassHalf \ Adversarial Analysis, fill=blue!10
      [
      \faGlasses \ Content Generation, fill=purple!10]
      ]
      ]
    ]
    [\textbf{Investigation} \\ Section \ref{sec:inv_opportunities}, fill=gray!5, align=center
      [\faPlay \ Transcription \\
        \faPlay \ Translation \\
        \faPlay \ Lead Generation \\
        \faPlay \ Document Synthesis \\
        \faPlay \ Image Analysis \\
        \faPlay \ Assisted Search \\
        \faPlay \ Trend Analysis, fill=green!10
        [
        \faGlasses \ Domain Research \\
        \faGlasses \ Synthetic Content \\ Detection \\
        \faGlasses \ Origin Identification \\
        \faGlasses \ Retrieval-Augmented \\ Fact-Checking, fill=purple!10]
        ]
    ]
    [\textbf{Audience Management} \\ Section \ref{sec:aud_opportunities}, fill=gray!5, align=center
        [\faPlay \ Metadata Generation \\
        \faPlay \ Format Conversion, fill=green!10
        [\faHourglassHalf \ Story Translation \\
        \faHourglassHalf \ Conversational Tipline \\
        \faHourglassHalf \ Live Fact-Checking, fill=blue!10
        [
        \faGlasses \ Web Page Generation \\
        \faGlasses \ Audience Analysis, fill=purple!10]
        ]
        ]
    ]
    [\textbf{Technology} \\ Section \ref{sec:tech_opportunities}, fill=gray!5, align=center
        [\faPlay \ Data Preprocessing \\
        \faPlay \ Data Summarization \\
        \faPlay \ Core NLP Tasks, fill=green!10
        [
        \faHourglassHalf \ Handling Multimodal \\ Misinformation \\
        \faHourglassHalf \ Addressing Data Scarcity, fill=blue!10
        [
        \faGlasses \ Detecting Coordinated \\ Inauthentic Behavior, fill=purple!10
        ]
        ]
        ]
    ]
    [\textbf{Advocacy} \\ Section \ref{sec:adv_opportunities}, fill=gray!5, align=center
        [
        \faGlasses \ Connecting Fact-\\Checking Organizations \\
        \faGlasses \ Promoting \\ Information Literacy \\
        \faGlasses \ Changing the Tenor \\ of Fact-Checking, fill=purple!10]
    ]
  ]
\end{forest}

\hfill \break

\begin{tabular}{|c|c|c|}
\toprule
\multicolumn{3}{|c|}{\cellcolor{gray!5}\textbf{Maturity}} \\
\midrule
\cellcolor{green!10}\faPlay \ In Use & \cellcolor{blue!10}\faHourglassHalf \ In Progress & \cellcolor{purple!10}\faGlasses \ Envisioned \\
\bottomrule
\end{tabular}

\caption{\footnotesize A description of In Use, In Progress, and Envisioned generative technologies grouped according to five fact-checking infrastructures.}
    \label{fig:opportunities_tree}

\end{figure*}

\section{Findings: Opportunities and Challenges of Generative AI in Fact-Checking}

\subsection{RQ1: Opportunities of Generative AI}

We found during thematic analysis that the technologies used and envisioned by fact-checking organizations depend largely on the organizational infrastructures into which they would be integrated. These infrastructures accorded to a large degree with the work of \citet{juneja2022human}, which described the sociotechnical work of editors; fact-checkers; social media managers; and long-term advocators. Drawing on these roles, we organize our findings according to the following divisions of organizational infrastructure, as shown in Fig. \ref{fig:opportunities_tree}: Editing, Investigation, Audience Management, Advocacy, and Technology, the last of which is new to our work but necessary to describe the impact of generative AI on the work of software developers and data scientists who build and maintain the data science pipelines employed by fact-checking organizations. \looseness=-1

When describing a technology, we also identify its status according to three levels of maturity, denoted using icons:

\begin{itemize}[left=1mm]
    \item \textbf{\faPlay \ In Use:} Technologies presently in use by participants, denoted with a rightward arrow to evoke a ``Play'' symbol.
    \item \textbf{\faHourglassHalf \ In Progress:} Technologies undergoing prototyping, testing, betas, or development by our participants, denoted with an hourglass symbol to communicate that some time remains before these technologies will be implemented.
    \item \textbf{\faGlasses \ Envisioned:} Technologies envisioned but not implemented or prototyped by our participants, denoted with eyeglasses to communicate that these technologies are further away and not yet in development.
\end{itemize}

\noindent Where fact-checking organizations reported achieving differing levels of maturity for a technology, we describe the most mature, and make note if this level of maturity has not been achieved by most other fact-checking organizations.

\subsubsection{\textbf{Editing}}\label{sec:ed_opportunities}

Editing ensures fact-checking content is engaging, approachable, and error-free, and spans from the beginning of a fact-check to publication, as editors are often involved in deciding which claims are worthy of a fact-check \cite{juneja2022human}. Participants in our study who were involved in Editing usually reported managing small teams. \looseness=-1

Many participants described using generative AI to refine and restructure fact-checks and internal reports, which then undergo human review. P3 reported using ``premium'' ChatGPT for \faPlay \ \textbf{Content Formatting} - to edit and refine written reports, and to restructure ``dense content'' into an approachable format for readers. P18 noted using ChatGPT to help in ``brushing up text.'' Several participants described in-development applications of generative AI for systematic \faHourglassHalf \ \textbf{Quality Assurance}, to prevent cosmetic and substantive editorial errors. P35 reported using ChatGPT to highlight grammatical and factual errors, tasks for which they normally use Grammarly \cite{grammarly}. P27 developed an app to address such errors:

\blockquote{\small We began with simple mistakes, geographical errors, or misspellings [...] I made a little Shiny app out of [ChatGPT] and showed it around, and people were really like [...] I can really see how this helps us. [...] And my aim is that we don't have these kind of simple mistakes anymore [...] this would be a huge achievement because simple mistakes are trivial on the one hand, but on the other hand [...] it's really important for the sense of quality the reader has and for trust.\looseness=-1}

\noindent P3 expanded on this view and envisioned generative AI providing \faHourglassHalf \ \textbf{Adversarial Analysis} before publishing a fact-check, referencing a strategy in development at a Sudanese newsroom: ``They use generative AI to actually give it an article that has been written, and ask the model to actually tell us whether there are any assumptions that have been met, that are inaccurate or incorrect.'' Finally, P8 envisioned fine-tuning a generative model on verified fact-checking articles, and using it for \faGlasses \ \textbf{Content Generation} at the beginning of the composition process, improving ``fact-checking output [by] pre-writing those fact-check articles,'' provided that the content did not require deep scientific knowledge of a topic, and remained subject to human review. While some participants envisioned using generative AI for content generation, many expressed deep discomfort with generative AI writing fact-checks, a finding explored in sections \ref{sec:tech_challenges} and \ref{sec:value_tensions}. \looseness=-1

\subsubsection{\textbf{Investigation}}\label{sec:inv_opportunities}

Investigation refers to the process of assessing the accuracy of potentially misinforming claims, and involves monitoring online sources for misinforming content; gathering verified sources to substantiate or refute claims; and writing a fact-check or internal report \cite{juneja2022human}. 19 of our 38 participants described working primarily in Investigation. \looseness=-1

Participants described many ongoing and envisioned uses of generative AI to perform tasks related to investigation and research assistance. As P25 noted, generative AI is used in common and ``taken-for-granted'' tasks like \faPlay \ \textbf{Transcription}, which save time and money for fact-checkers. P20 highlighted the use of generative AI for \faPlay \ \textbf{Translation} of internet content in need of investigation, such as ``many translations from Ukraine'' due to misinformation related to the Russia-Ukraine war. Participants also described adopting generative AI to directly support investigations. P11 noted that, while they never use ChatGPT for writing fact-checks, they use it for \faPlay \ \textbf{Lead Generation}, ``trying to generate ideas for stories.'' P28 used ChatGPT for \faPlay \ \textbf{Document Synthesis}, to save time by organizing research notes and summarizing text from web pages. P19 used GPT-4 for \faPlay \ \textbf{Image Analysis}, substituting the model for reverse image search in some cases, noting they can ``ask it where the photo was taken, and sometimes we [get the] correct answer,'' or useful hints for continuing the search. P25 fine-tuned GPT-3.5-Turbo to perform \faPlay \ \textbf{Assisted Search}, generating custom Google search queries, often in a language not spoken by the fact-checker, noting that such a task would ``take me hours to do and I still might miss some of the terms.'' P14 reported using ChatGPT for \faPlay \ \textbf{Trend Analysis}, to keep abreast of media produced by websites known to produce misinforming content: ``We take the top 200 headlines from the last 24 hours from those sites [...] and run them through ChatGPT, asking it to summarize the main narratives [...] and extract the names of people, places, entities [...] and then send that to me by email. So every six hours, [...] we get an email.''

Participants envisioned technologies to increase their expertise and verify novel content. P29 noted fact-checkers need to ``become a little mini expert in a certain specific topic,'' and envisioned a technology for \faGlasses \ \textbf{Domain Research}, summarizing and collating literature for review. P24 envisioned using generative AI for \faGlasses \ \textbf{Synthetic Content Detection}, identifying content produced by AI. P24 described this as their ``holy grail'' given recent increases in synthetic content and the difficulty of fact-checking it. P29 envisioned a tool for \faGlasses \ \textbf{Origin Identification}, scanning the internet for the first occurrence of content, bringing fact-checkers ``closer to the verification.'' Finally, P26 envisioned a \faGlasses \ \textbf{Retrieval-Augmented Fact-Checking} system that could ``retrieve data in almost real time, to consult with databases.''

\subsubsection{\textbf{Audience Management}}\label{sec:aud_opportunities}

Audience management refers to processes supporting the publication and wide dissemination of fact-checking content. Audience managers increase engagement by employing SEO optimization, online advertising, and conversion of written content to short videos \cite{juneja2022human}. Our interviews revealed that audience management involves connecting with consumers of fact-checks over many channels, of which social media is one.

Participants used generative AI to both connect with existing audiences and reach new audiences. P33 used generative AI for \faPlay \ \textbf{Metadata Generation} to support social media content, including ``summaries, SEO for article publishing, title generation.'' P14 used generative text-to-speech models for \faPlay \ \textbf{Format Conversion}, taking fact-checks and converting them into audio for short videos posted to ``Tiktok, Instagram, YouTube shorts,'' noting that AI helps achieve the right volume for disseminating content via video sharing algorithms. P31 described working on AI-based \faHourglassHalf \ \textbf{Story Translation} of their fact-checked content into multiple languages, a goal echoed by P14, who envisioned translating their short videos.

Participants also described efforts to connect immediately with audiences, before misinformation could go viral. P32 described a beta of a fact-checking chatbot in a \faHourglassHalf \ \textbf{Conversational Tipline} using OpenAI's GPT-4, from which they collect circulating misinformation from users and instantaneously deliver information to audiences who use more private platforms like WhatsApp, rather than Facebook and X. Other participants, including P1 and P17, described in-progress chat-based tiplines via which users can submit suspected misinformation. P13 described an in-progress tool for \faHourglassHalf \ \textbf{Live Fact-Checking} that ``can do claim matching while a person is speaking,'' allowing for claims to be debunked in real time.

Participants envisioned tools for automatic web content generation and predictive analysis of audience engagement. P28 envisioned automatic \faGlasses \ \textbf{Web Page Generation} that could produce fact-checks ``based on social media posts that are verified [...] and then just code the iFrames for us to be able to embed it in our own content,'' saving programming labor. Finally, P12 envisioned a system for \faGlasses \ \textbf{Audience Analysis}, providing insight into how audiences would consume a factcheck, and recommending it be presented as a video, an infographic, or a short or long-form article.

\subsubsection{\textbf{Technology}}\label{sec:tech_opportunities}

Technology refers to work building and maintaining data science pipelines used by fact-checking organizations. While not all organizations have a Technology unit, many uses of generative AI would be invisible without specific reference to the work of software engineers and data scientists employed by fact-checkers and their partners.

Participants described in-progress generative technologies to improve the core functionality and end user experience of fact-checking data science pipelines. P2 used generative AI for \faPlay \ \textbf{Data Preprocessing}, to ``get a rewriting or a restatement of the claim that's a bit cleaner'' than unprocessed social media content or tipline messages. P8 noted that generative models improve on \faPlay \ \textbf{Core NLP Tasks} over finetuned BERT models - ``the previous generation'' of NLP - including for claim content matching. \faPlay \ \textbf{Data Summarization} for human end users was another predominant use reported by our participants. For example, P2 described using generative AI to provide a human-readable summary of clusters of misinforming content, describing ``the variety of content'' and ``how it's changing over time.'' P25 tested the capability of Google's generative models for natively \faHourglassHalf \ \textbf{Handling Multimodal Misinformation}, wherein the relationship between text and image must be parsed to understand subtle misinformation or hate speech. P2 described ongoing experiments for \faHourglassHalf \ \textbf{Addressing Data Scarcity}, noting that they would ``generate pseudo labeled data'' where real, ``gold standard'' data for novel misinformation did not exist, and either ``use those labels directly or use them to train a lower cost classifier.'' Finally, P16 envisioned a generative model for \faGlasses \ \textbf{Detecting Coordinated Inauthentic Behavior} and influence operations, noting AI might ``do more meaningful work on [detecting] coordinated networks, behavior.''

\subsubsection{\textbf{Advocacy}}\label{sec:adv_opportunities}

Advocacy refers to long-term processes to influence information policy, forge connections between fact-checking organizations, and engage with the public via  misinformation literacy campaigns \cite{juneja2022human}. Participants performing advocacy work were often senior managers who also managed teams of investigators.

Participants suggested generative AI could encourage information literacy, promote relationships between organizations, and improve access to information. P8 envisioned generative AI helping in \faGlasses \ \textbf{Connecting Fact-Checking Organizations}, by standardizing the methods and technologies to combat misinformation in Europe, noting that recent models handle most European languages well. P12 envisioned generative AI scaling fact-check operations by connecting organizations across Africa: ``I think that's one area in which generative AI can really help. If fact-checkers are working together [...] they can help scale the impact of their fact check to different segmented audiences that they serve [...], whether it's local language, whatever format.'' P21 envisioned generative AI for \faGlasses \ \textbf{Promoting Information Literacy}: ``people will have the option to kind of play games with the chatbot that are intended for media literacy on misinformation.'' Finally, P30 envisioned generative AI \faGlasses \ \textbf{Changing the Tenor of Fact-Checking}, shifting the way audiences consume fact-checking content: \looseness=-1

\blockquote{\small Changing the way users can consume reliable and good information could be incredibly beneficial for fact-checkers [...] If [...] they need and get good information [...] with a chatbot, for example, or with other ways, that would be fantastic. [...] people value us as we are because we give them reliable information and they know they can trust us, but if they also knew that they can consume [that] information in any way they wanted to, I think it would be an incredible leap forward. \looseness=-1}

\subsection{RQ2: Challenges and Limitations}

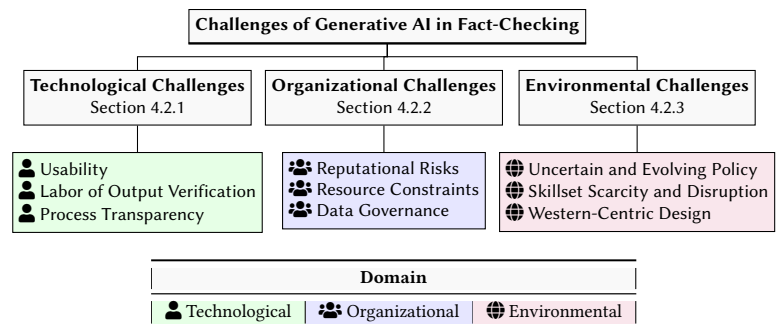
\begin{figure*}
\footnotesize
\centering
\begin{forest}
  for tree={
    align=left,
    font=\sffamily,
    l sep'=10pt,
    s sep = 5pt,
    fork sep'=5pt,
    draw,
  },
  forked edges,
  baseline,
  where level=0{
    tikz={\draw [thick] (.children first) -- (.children last);},
  }{},
  [\textbf{Challenges of Generative AI in Fact-Checking}, fill=gray!5
    [\textbf{Technological Challenges} \\ Section \ref{sec:tech_challenges}, fill=gray!5, align=center
        [\faUser \ Usability \\
        \faUser \ Labor of Output Verification \\
        \faUser \ Process Transparency, fill=green!10
    ]    
    ]
    [\textbf{Organizational Challenges} \\ Section \ref{sec:org_challenges}, fill=gray!5, align=center
        [\faUsers \ Reputational Risks \\
        \faUsers \ Resource Constraints \\
        \faUsers \ Data Governance, fill=blue!10
        ]
    ]
    [\textbf{Environmental Challenges} \\ Section \ref{sec:env_challenges}, fill=gray!5, align=center
        [\faGlobe \ Uncertain and Evolving Policy \\
        \faGlobe \ Skillset Scarcity and Disruption \\
        \faGlobe \ Western-Centric Design, fill=purple!10
    ]
  ]
  ]
\end{forest}

\hfill \break

\begin{tabular}{|c|c|c|}
\toprule
\multicolumn{3}{|c|}{\cellcolor{gray!5}\textbf{Domain}} \\
\midrule
\cellcolor{green!10}\faUser \ Technological & \cellcolor{blue!10}\faUsers \ Organizational & \cellcolor{purple!10}\faGlobe \ Environmental \\
\bottomrule
\end{tabular}
    \caption{\footnotesize A description of the Technological, Organizational, and Environmental challenges to generative AI in fact-checking}
    \label{fig:challenges_tree}

\end{figure*}

We found during thematic analysis that, unlike RQ1, challenges did not break down based on organizational infrastructure. Rather, participants described challenges related to using the technology itself; to incorporating it in an organization; and to factors that affected society as a whole, and were often out of their organization's control. As illustrated in Fig. \ref{fig:challenges_tree}, the Technology-Organization-Environment (TOE) framework \cite{prasad2023towards} offers a ready model for these findings, and we use it to describe participant challenges as follows: \faUser \ \textbf{Technological Challenges} that impact the user of a system, such as the manual labor of verifying generative model outputs, denoted with an icon of a person (note that Technological challenges are in fact sociotechnical, involving human interaction with technology \cite{yu2023antecedents}); \faUsers \ \textbf{Organizational Challenges} that impact an organization in the aggregate, such as reputational risks incurred by using systems that hallucinate, denoted with an icon of multiple people; and \faGlobe \ \textbf{Environmental Challenges} that impact not only an organization but an entire society, such as the scarcity of skilled workers, denoted with a globe icon. 

\subsubsection{\faUser \ \textbf{Technological Challenges}}\label{sec:tech_challenges}

Participants described barriers related to model usability; the labor of verifying model output; and the conflict in using a hard-to-explain technology in a process requiring absolute transparency.

\noindent \faUser \ \textbf{Usability}: Participants noted a lack of clarity concerning prompt engineering and hyperparameter tuning. P27 described an iterative process of choosing prompts for OpenAI models that resulted in uncertainty: ``We prompted and we coded a bit and we thought, oh, this prompting technique, and combining this prompt with that one and iterating it, and then majority rule. And we [...] thought, okay, is this really the way that this should be used?'' P2, who develops AI for fact-checking, noted that generative AI is ``incredibly sensitive'' to prompts, and ``I'm sure we don't have the best approach'' to prompt design. P8 noted the verbosity of ChatGPT reduced its usefulness for fact-checking content, which is ``laser-focused.'' P14 noted that OpenAI models used for summarization randomly enter ``loops'' of repeating one word. P25 tweaked settings like temperature to improve reliability, but the effects were hard to see in model output. Finally, P13 expressed frustration over failures of image models like DALL-E to render text in images, limiting their use in creating visual content for stories. \looseness=-1

\noindent \faUser \ \textbf{Labor of Output Verification}: Every participant described human review of AI-generated content as non-negotiable for ensuring the quality of published fact-checking content. Participants described some uses of generative AI as currently untenable due to the verification labor required. P11 summed up participants' opinions: ``Of course it's not as accurate. The tools are not as accurate. You still need to corroborate the information that you get.'' P25 noted that while it is ``tempting to use [generative AI] for speeding up your work,'' its unreliability means fact-checkers must ``see [if] this is correct, what is the source?'' P36 noted that, even if ChatGPT provides a lead or answers a question, ``it's just as quick for us to go and find it [...] we're just so used to that lateral sort of work. And to be honest [...] we'd be going and double checking all that anyway.'' P14 noted that, like the internet before it, generative AI re-organized the efforts of fact-checkers to fit the technology, noting ``we had to train them'' on writing styles that led to better outcomes with generative AI. \looseness=-1

\noindent \faUser \ \textbf{Process Transparency}: Participants described generative AI as a potential impediment to the transparency needed to create trustworthy content. P29 noted, ``Our sourcing is [...] always actually quite transparent. [...] we fill our story with hyperlinking to our sources and [...] how we got to everything.'' P8 noted that hallucinations prevented them from using ChatGPT: ``The result was largely unusable [...] The sources have to be very well integrated [...] it just doesn't work. Sounds very good, [but] there will be hallucinations, it will just make up sources.'' P12 said explaining research that uses generative AI is hard because ``as fact-checkers, we actually do not understand the processes'' of the models. Finally, P35 noted that generative AI may engender questions about bias concerning selection of experts: ``Have [models] weighed whether [...] there are uncertainties about them, have they been disgraced for some reason? [...] humans have biases as well, but I think in factcheck [...] there's always many, many different sets of eyes on our checks and the experts we use.''

\subsubsection{\faUsers \ \textbf{Organizational Challenges}}\label{sec:org_challenges}

Participants described organizational barriers including reputational risks in unpredictable technology; resource constraints preventing investment; and concerns over data provenance and ownership.

\noindent \faUsers \ \textbf{Reputational Risks}: The most common organizational barrier participants identified was the reputational risk of a mistake in a fact-check. P16 said ``just by default we need to be much more cautious than almost anyone else, because it's hard for us to come back from a big mistake.'' P8 and P26 both noted that 90\% accuracy is insufficient for fact-checkers, whose relationship with audiences depends on offering information verified by experts. P2 noted that sharing generative technologies with partner organizations also shares risk, prompting further caution: ``It's their organization's reputation that's at risk, not ours only.'' P18 said they would not trust generative models when there is only one correct answer. Finally, P2 noted that academic evaluations of generative AI are unreliable, as fact-checkers handle \textit{novel} information: ``In an academic context, it's always retrospective. [...] you put that into the Bing API or Google and you find lots of relevant content that can help refute that claim. But when it first appeared [...] I don't think that was the case.'' 

\begin{figure*}[htbp]
\centering
\begin{tikzpicture}

  \draw[->] (0,0) -- (6,0) node[right, align=center] {\textbf{Human} \\ \textbf{Producer}};
  \draw[->] (0,0) -- (-5.55,0) node[left, align=center] {\textbf{Generative} \\ \textbf{AI Producer}};

  \draw[->] (0,0) -- (0,2) node[above] {\textbf{Human Verifier}};
  \draw[->] (0,0) -- (0,-2) node[below] {\textbf{Generative AI Verifier}};

  \node[right, draw, fill=green!10, minimum width=7.25cm, align=center, rounded corners] at (.125, .75) {\faPlay \ \textbf{Editorial Review} of human-written content};
  \node[right, draw, fill=green!10, minimum width=7.25cm, align=center, rounded corners] at (.125, 1.5) {\faPlay \ \textbf{Managerial Review} of junior employee work};
  
  \node[right, draw, fill=blue!10, minimum width=7.25cm, align=center, rounded corners] at (.125, -.75) {\faHourglassHalf \ \textbf{Quality Assurance} tools for catching editorial errors};
  \node[right, draw, fill=blue!10, minimum width=7.25cm, align=center, rounded corners] at (.125, -1.5) {\faHourglassHalf \ \textbf{Adversarial Analysis} of human-written content};

  \node[left, draw, fill=green!10, minimum width=7.25cm, align=center, rounded corners] at (-.125, -.75) {Automatic prompt refinement for \faPlay \ \textbf{Core NLP Tasks}};
  \node[left, draw, fill=blue!10, minimum width=7.25cm, align=center, rounded corners] at (-.125, -1.5) {Synthetic data for \faHourglassHalf \ \textbf{Addressing Data Scarcity}};
  
  \node[left, draw, fill=purple!10, minimum width=7.25cm, align=center, rounded corners] at (-.125, .75) {\faGlasses \ \textbf{Content Generation} with finetuned models};
  \node[left, draw, fill=green!10, minimum width=7.25cm, align=center, rounded corners] at (-.125, 1.5) {\faPlay \ \textbf{Metadata Generation} for social media content};

\end{tikzpicture}

\footnotesize
\begin{tabular}{|c||c|c|}
\toprule
\multicolumn{3}{|c|}{\cellcolor{gray!5}\textbf{Maturity}} \\
\midrule
\cellcolor{green!10}\faPlay \ In Use & \cellcolor{blue!10}\faHourglassHalf \ In Progress & \cellcolor{purple!10}\faGlasses \ Envisioned \\
\bottomrule
\end{tabular}
\caption{Designing for verification: A sociotechnical verification space for the production and verification of content.}
\label{fig:verification_space}
\end{figure*}
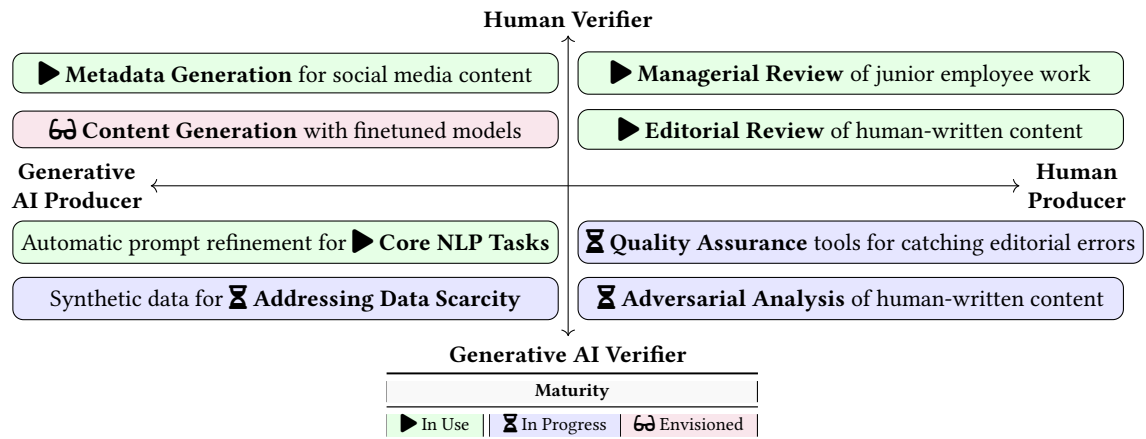

\noindent \faUsers \ \textbf{Resource Constraints}: Participants said that most fact-checking organizations lack the financial resources to invest in generative AI. P3 noted that donor funding informs the building of new technologies: ``We rely on donor funding a lot, and donor funding is to address a specific use case. So if there's not enough resources here marked for building a machine learning model, then, we just do that out of pocket or partner with other organizations. So that has been a main limitation for us.'' P27 notes that, even at their well-resourced organization, ``we don't have really this AI development department'' and that colleagues in low-resource organizations only develop AI tools with universities. P13 said that their organization cannot afford tools developed by better-resourced fact-checkers: ``[I] met the team of <Organization> at the last IFCN conference, and they told me it's going to be a huge sum to get that subscription [...] a small company with seven people, [we] might not be able to afford that.'' P21 noted resource constraints facing organizations in the global south: 

\blockquote{\small We're in the global south, so sometimes the resources are not the same, either to use generative AI [...] for investigating or creating our own tools. For example, some colleagues in Spain have a chatbot [...] and right now we're trying to find resources to buy the chatbot, the basic form of the chatbot [...] it's more like an economic problem and it's not exclusive for [us], but probably more small fact-checking organizations in different global south countries. \looseness=-1}

\noindent \faUsers \ \textbf{Data Governance}: Participants expressed concerns about the privacy of their data and the provenance of AI training data. P27 noted using open source models when possible, as they ``have very sensitive material [...] investigative reporting and investigative stories, and we don't want this to be used in models and as training material.'' P27 appreciated the ``legal safety'' of European data protection laws, recalling a conversation with a fact-checker who highlighted the importance of trust between organizations: ``I really would like to work with all these Google tools, but still, it's Google and I'm kind of hesitating. I wish the New York Times would've developed it. Then it would be very easy for me to trust it.'' P31 noted many organizations questioned if they should be compensated for the labor of producing content used to train generative models: ``You'll just continue to see more interest in using fact-check information to feed AI [...] Are we going to be compensated?'' P12 described the uncertainty of what data was used to train models as ``problematic'' for fact-checkers. \looseness=-1

\subsubsection{\faGlobe \textbf{Environmental Challenges}}\label{sec:env_challenges} Participants described society-wide barriers including uncertain government and partner policy; skillset scarcity and disruption; and western-centric design. 

\noindent \faGlobe \ \textbf{Uncertain and Evolving Policy}: P16 contended that, though evolving, government policies in Europe are not equipped to deal with generative AI, as well as the new forms of misinformation arising from it. P13 noted that law around generative AI in India is limited by technicalities and intended primarily to stop citizens from being ``cheated'' by deepfakes. P26 expressed concern about ``impinging on the freedom of speech'' if generative AI were overapplied for moderating speech, including that produced by AI models. P8 noted uncertainty about the policies of networks like the IFCN, saying ``I think you can't just use that too much in your work, if you want to stay within the framework, which is super important for us.''

\noindent \faGlobe \ \textbf{Skillset Scarcity and Disruption}: Participants consistently noted the short supply of generative AI skills. P35 said this scarcity rendered them unaffordable: ``people who do know how to do that are working in organizations where they're on a much higher wage than anything that a fact-checking or journalism organization could offer.'' P3 said finding tech workers in Africa with generative AI skills was difficult and made harder by the headhunting practices of U.S. tech companies, noting ``getting the right skill at a level we can actually afford as an NGO has been a [...] major challenge.''

Many participants reported attempts to build generative AI skills internally, in part to prepare for its disruptive impact on fact-checking workflows. P25 described generative AI skills as important both for efficiently dealing with misinformation and for awareness of the misinforming content that generative AI enables. P27 noted difficulty incorporating generative AI due to fears of displacement by the technology, and of changes to the fact-checking profession: ``I think the biggest challenge is how to communicate this in a department [...] It won't replace us, probably, but it will change our work tremendously.'' P28 characterized generative AI as a generational issue for local fact-checkers, noting that aversion to new technology hampers the ability to match promulgators of misinformation: ``We are not even yet at the point of being comfortable appearing on videos [...] if you [...] confidently face your cell phone and record what is happening, the same way malign actors are, maybe we would have a fighting chance. And I'm not even at the point of talking about AI yet.'' \looseness=-1

\noindent \faGlobe \ \textbf{Western-Centric Design}: P2, P4, and P11 each said that, while generative AI tools perform well in English, their quality remains poor for low-resource languages, particularly local African languages. P13 described the local languages they dealt with as a low priority for companies developing language models. P12 highlighted the need for Africa-centered partnerships to overcome western biases and lack of access to the data for training African models:

\blockquote{\small The challenge we have right now is, it's like the new shiny toy and everybody is talking about AI in Africa, but when you actually ask from an African point of view [...], the problem comes with [...] integrated bias in Eurocentric, or let me just say American platforms. So Meta, Google, they're all developed there and then they're used here. So the inherent biases in that, that's not something that we can do anything about [...] it'll be good to find ways, at least for African tech, to partner with these people, to develop the tools that would work for the continent. Otherwise, it's just noise.\looseness=-1}

\noindent P28 connects AI language with inauthentic colonial power: ``In post-colonial places like ours, we are taught to read in another language. It's monotonous. And what I'm getting at is that that monotony is the same sound you hear in AI, and it appears very authentic, even if it's not [...] It really affects the way we are being shaped as a country.'' \looseness=-1

\begin{table*}[t]
    \centering
    \footnotesize
    \begin{tabular}{|C{1.7cm}|L{7.35cm}|C{.9cm}|L{3.75cm}|}
    \toprule
    \multicolumn{1}{|c|}{IFCN Principle} & \multicolumn{1}{|c|}{IFCN Description} & \multicolumn{1}{|c|}{Value} & \multicolumn{1}{|c|}{Tension with Generative AI} \\
    \midrule
     Non-Partisanship and Fairness & ``Signatory organizations fact-check claims using the same standard for every fact-check. They do not concentrate their fact-checking on any one side. They follow the same process for every fact-check and let the evidence dictate the conclusions. Signatories do not advocate or take policy positions on the issues they fact-check.'' & Fairness & Generative AI exhibits wide variance depending on the structure of a user's input, and may reflect both implicit and explicit societal biases based on its training and fine-tuning data. \\
     \hline
     Standards and Transparency of Sources & ``Signatories want their readers to be able to verify findings themselves. Signatories provide all sources in enough detail that readers can replicate their work, except in cases where a source’s personal security could be compromised. In such cases, signatories provide as much detail as possible.'' & Transparency & Models cannot affirmatively identify sources, and may hallucinate inaccurate sources of information when asked to do so. \\
     \hline
     Transparency of Funding and Organization & ``Signatory organizations are transparent about their funding sources. If they accept funding from other organizations, they ensure that funders have no influence over the conclusions the fact-checkers reach in their reports. Signatory organizations detail the professional background of all key figures in the organization and explain the organizational structure and legal status. Signatories clearly indicate a way for readers to communicate with them.'' & Accountability & Models are pretrained using poorly specified data, and may be aligned via practices exploiting low-cost workers in developing countries. Developers deny responsibility to compensate producers of web-scraped data.   \\
     \hline
     Standards and Transparency of Methodology & ``Signatories explain the methodology they use to select, research, write, edit, publish and correct their fact-checks. They encourage readers to send claims to fact-check and are transparent on why and how they fact-check.'' & Explainability & Even when models produce a correct answer, they cannot give a reliable explanation of how it was arrived at. \\
     \hline
     Open and Honest Corrections Policy & ``Signatories publish their corrections policy and follow it scrupulously. They correct clearly and transparently in line with the corrections policy, seeking so far as possible to ensure that readers see the corrected version.'' & Openness & Generative AI opens a channel to disseminate information not easily observed or corrected by experts. \\
     \bottomrule
    \end{tabular}
    \caption{\footnotesize \centering We leverage the IFCN principles to identify fact-checking values, and participant insights to describe tensions with generative AI.}
    \label{tab:value_tensions}
\end{table*}

\section{Discussion}

Realizing opportunities of generative AI in fact-checking requires not only building new technical competencies, but also addressing responsible use in a domain concerned primarily with reliability. We introduce a novel sociotechnical dimension to the design space for generative models that centers  information verification; discuss tensions between the values of fact-checking and the values of generative AI; and outline a research agenda for generative AI in fact-checking. \looseness=-1

\subsection{Design Considerations for Generative AI in Fact-Checking}

\subsubsection{Designing for Verification}

Every participant centered one concern about generative AI: verification. This arose in Organizational challenges, in reputational risks of publishing unverified output, and in Technological challenges, as fact-checkers must transparently explain their processes and verify model output.

To begin addressing these challenges, we introduce a sociotechnical dimension to the design space of generative models focused on the production and verification of content, conceived using a 2x2 matrix. We locate the \emph{Producer} of content on the X axis, and the \emph{Verifier} of content on the Y axis (see Fig. \ref{fig:verification_space}). The upper right (\emph{Human Producer, Human Verifier}) characterizes workflows in most fact-checking and digital media companies, as experienced staff review content authored by junior staff, and editors refine content authored by fact-checkers and writers. The bottom right (\emph{Human Producer, GenAI Verifier}) characterizes the \faHourglassHalf \ \textbf{Quality Assurance} system for catching small errors (P27), and the \faHourglassHalf \ \textbf{Adversarial Analysis} of fact-checking content (P3). It adds security for high-stakes tasks like fact-check publication that are performed primarily by humans. Fact-checkers generally agreed that the bottom left quadrant (\emph{GenAI Producer, GenAI Verifier}), which includes no human oversight, is suitable for low-stakes settings, or where there is a clear evaluation metric that can be used by the verifying model. For example, P2 used generative AI to refine prompts used by other generative models in \faPlay \ \textbf{Core NLP Tasks}, improving pipeline performance with little human oversight. Finally, participants held mixed views of the upper left quadrant (\emph{GenAI Producer, Human Verifier}), noting that the \faUser \ \textbf{Labor of Output Verification} may render such designs inefficient, or devalue the role of the human. Participants welcomed AI for \faPlay \ \textbf{Metadata Generation} and \faPlay \ \textbf{Format Conversion} of content into new modalities, noting such uses saved time, even with human review. \looseness=-1

Verification is essential for domains where generative AI may provide transformative benefits, but where the consequences of incorrect output are high. Consider applications of generative AI in law \cite{chatlaw}, where high-profile mistakes have rendered use of generative models suspect, or medicine, where research suggests demographic biases emerge in models trained on medical text \cite{Adam2022WriteIL}. The verification dimension, envisioned for the high-stakes, information-centered domain of fact-checking, provides a framework for conceptualizing applications that value veracity at least as highly as efficiency. \looseness=-1

\subsubsection{Describing Value Tensions}\label{sec:value_tensions}

Value tensions refer to cases wherein the values of a stakeholder in a technology or process come into conflict with the values of another stakeholder or of the technology itself \cite{10.1561/1100000015}. Drawing on \citet{birhane2022values}, we do not view generative AI as value-neutral by default, and we conduct a conceptual investigation into the tensions between the values of fact-checking in interaction with generative AI \cite{wolk2007}. We use the \href{https://ifcncodeofprinciples.poynter.org/know-more/the-commitments-of-the-code-of-principles}{five principles} of the IFCN, an organization founded to promote common standards in fact-checking \cite{ifcn}, as a basis for examining the values of fact-checking. Building on insights from our interviews with IFCN signatories and partners, we map each IFCN principle to an underlying value, and describe its tension with generative AI. Table \ref{tab:value_tensions} describes these values and value tensions in detail, illustrating conflicts between generative AI and core fact-checking values such as fairness, transparency, and explainability. While many primarily technical tensions are resolvable, especially in the context of human-AI collaboration, tensions surrounding values like Accountability may require significant changes in the relationships between information professionals like fact-checkers and the technology companies that benefit from their labor \cite{10.1145/3593013.3594070}.

\begin{table*}[]
    \centering
    \small
    \begin{tabular}{|C{2.5cm}|L{10cm}|C{1.5cm}|}
    \toprule
    \multicolumn{1}{|c|}{\textbf{Fairness}} & Directions that seek to mitigate both technical and societal bias and unfairness. & Interest \\
    \hline
    \hline
    Technology for the Global South & Building technologies in coordination with fact-checking organizations in the global south to improve model performance and usability, especially in local languages. & P2, 3, 11, 12, 13, 28 \\
    \hline
    Detecting and Mitigating Bias & Developing technical and human-centered approaches to identifying and minimizing bias in model output and human-AI collaborations. & P5, 6, 21, 33, 35 \\
    \hline
    Combating Information Inequality & Developing methods to reach audiences outside of well-educated, well-resourced communities that typically consume fact-checking content. & P4, 9, 17, 29, 30, 31 \\
    \hline
    \hline
    \multicolumn{1}{|c|}{\textbf{Accountability}} & Directions to improve accountability of AI developers to users, fact-checkers to audiences. & Interest \\
    \hline
    \hline
    Improving Data Standards and Safety & Developing technical and policy approaches to ensuring that fact-checker data and content is not misused when training or fine-tuning generative AI models. & P27, 29, 31, 37 \\
    \hline
    Auditing for Deceptive Design & Auditing generative AI systems for deceptive design patterns that manipulate human users into placing too much faith in the veracity of their output. & P1, 12, 25, 26, 28 \\
    \hline
    Improving Open Models & Developing highly usable open source and open weight generative AI models to alleviate fact-checker concerns related to the privacy and ownership of their data. & P2, 5, 14, 16, 27, 37 \\
    \hline
    \hline
    \multicolumn{1}{|c|}{\textbf{Transparency}} & Directions to equip fact-checkers with designs and approaches to maximize transparency. & Interest \\
    \hline
    \hline
    Benchmark Development & Developing benchmarks that measure language model performance in settings closer to the real-world scenarios faced by fact-checkers, accounting for the novelty of misinformation. & P2, 3, 16, 37, 38 \\
    \hline
    Synthetic Content Detection & Developing approachable generative AI tools for reliably detecting synthetic content, whatever the modality (such as text, image, audio, or video). &  P3, 10, 13, 15, 22, 23, 24 \\
    \hline
    Designing for Transparency & Developing design spaces and methodologies that center the transparent processes of fact-checking professionals and organizations. & P27, 31, 36 \\
    \hline
    \bottomrule
    \end{tabular}
    \caption{\footnotesize We propose nine research directions for generative AI in fact-checking in which our study participants expressed interest.}
    \label{tab:research_directions}
\end{table*}

\subsubsection{Defining A Research Agenda}

We describe directions for research identified by the participants of our interview study, grouped by the research community (fairness, accountability, or transparency) they primarily address (see Table \ref{tab:research_directions}). Participants stressed that researchers at universities and partner organizations will play an important role in advancing these research directions, but that research cannot have meaningful impact without the involvement of fact-checking organizations. P37 noted the limited effort of researchers to connect with fact-checkers, saying, ``normally researchers are doing research without even talking with fact-checkers, so they don't know how the fact-checking world works.'' Some directions, such as developing technology for the global south, or combating information inequality, may also require researchers to forge new relationships with individuals and organizations outside of their existing networks. \looseness=-1

\subsection{Limitations and Future Work}

Our findings are limited in that we focus solely on fact-checkers, and primarily on IFCN signatories and partners. There are many stakeholders in fact-checking, including audiences for fact-checks, technology providers, government bodies, and indirect beneficiaries of the impact of fact-checking on the information ecosystem \cite{juneja2022human}. Future work might center the interests and values of these stakeholders. Moreover, while we draw on the literature of organizational change, we are primarily concerned with understanding the evolution of organizations undertaking the work of information verification, rather than organizations broadly. Future work might seek to generalize or contextualize our findings with other organizations and sectors. \looseness=-1

\section{Conclusion}

We presented an interview study with $N$=38 participants at 29 fact-checking organizations across six continents, describing the opportunities and challenges of incorporating generative AI in sociotechnical fact-checking infrastructures. Insights from interviews formed the basis for a novel Verification dimension in the design space for generative models for fact-checking. The principles of the IFCN informed a description of the value tensions between fact-checking, which centers transparency, fairness, accountability, and reliability, and generative AI, an unpredictable and sometimes unreliable technology. Finally, we proposed a research agenda for generative technologies, designs, and approaches in fact-checking. \looseness=-1

\section*{Researcher Positionality}

The authors are not themselves fact-checkers, but academic researchers who approach the present work with a mix of sociological expertise related to fact-checking work and technical expertise with generative AI. The second author has maintained relationships with the global fact-checking community throughout their career. While we made concerted efforts to faithfully represent the views and experiences of fact-checkers operating globally, including in the global south, we nonetheless recognize that our positionality as academic researchers at a university in a developed country prohibits us from fully capturing the challenges faced and opportunities envisioned by information professionals operating in different contexts from ours. \looseness=-1

\section*{Ethical Considerations}

We note that our focus on IFCN signatories and their partner organizations may not be representative of all fact-checking organizations, including especially disadvantaged organizations with which we would not have interacted via a purposive and snowball sampling methodology. Moreover, we note that while fact-checkers play an essential role in maintaining the health of the information ecosystem in countries with well-established freedom of speech protections, their impact may be limited in less open environments for speech. As a study positioned within the context of a profession that addresses misinforming content, we thus note that we are not able to fully represent the efforts of marginalized individuals and organizations whose work nonetheless improves the health of the information ecosystem; our work extends only to those arenas in which fact-checkers operate. \looseness=-1

\section*{Adverse Impacts}

We considered the possibility that our work might elicit more enthusiasm than might be warranted with regard to adopting generative AI in human-centered discipline like fact-checking. While we have devoted large sections of the work to describing challenges, value tensions, and unrealized research, we note explicitly that this work should not be read as an endorsement of generative AI, but as qualitative scientific research describing the perspectives of a community around adopting this new technology. Cautious, human-centered design that centers the values of fact-checkers and the communities they serve will be necessary to ensure that generative AI is not employed for well-meaning but harmful applications in fact-checking, such as rendering the human labor of fact-checking invisible \cite{juneja2022human} or producing fact-checking content that ultimately lacks the authority of that of a human expert \cite{longoni2022news}, diluting the power of an essential mechanism for maintaining the health of information ecosystems. \looseness=-1

\bibliographystyle{ACM-Reference-Format}
\bibliography{references}


\begin{thebibliography}{92}


\ifx \showCODEN    \undefined \def \showCODEN     #1{\unskip}     \fi
\ifx \showDOI      \undefined \def \showDOI       #1{#1}\fi
\ifx \showISBNx    \undefined \def \showISBNx     #1{\unskip}     \fi
\ifx \showISBNxiii \undefined \def \showISBNxiii  #1{\unskip}     \fi
\ifx \showISSN     \undefined \def \showISSN      #1{\unskip}     \fi
\ifx \showLCCN     \undefined \def \showLCCN      #1{\unskip}     \fi
\ifx \shownote     \undefined \def \shownote      #1{#1}          \fi
\ifx \showarticletitle \undefined \def \showarticletitle #1{#1}   \fi
\ifx \showURL      \undefined \def \showURL       {\relax}        \fi
\providecommand\bibfield[2]{#2}
\providecommand\bibinfo[2]{#2}
\providecommand\natexlab[1]{#1}
\providecommand\showeprint[2][]{arXiv:#2}

\bibitem[Adam et~al\mbox{.}(2022)]%
        {Adam2022WriteIL}
\bibfield{author}{\bibinfo{person}{Hammaad Adam}, \bibinfo{person}{Ming Yang}, \bibinfo{person}{Kenrick~D. Cato}, \bibinfo{person}{Ioana Baldini}, \bibinfo{person}{Charles~R. Senteio}, \bibinfo{person}{Leo~Anthony Celi}, \bibinfo{person}{Jiaming Zeng}, \bibinfo{person}{Moninder Singh}, {and} \bibinfo{person}{Marzyeh Ghassemi}.} \bibinfo{year}{2022}\natexlab{}.
\newblock \showarticletitle{Write It Like You See It: Detectable Differences in Clinical Notes by Race Lead to Differential Model Recommendations}.
\newblock \bibinfo{journal}{\emph{Proceedings of the 2022 AAAI/ACM Conference on AI, Ethics, and Society}} (\bibinfo{year}{2022}).
\newblock
\urldef\tempurl%
\url{https://api.semanticscholar.org/CorpusID:248572183}
\showURL{%
\tempurl}


\bibitem[Bender et~al\mbox{.}(2021)]%
        {bender2021dangers}
\bibfield{author}{\bibinfo{person}{Emily~M Bender}, \bibinfo{person}{Timnit Gebru}, \bibinfo{person}{Angelina McMillan-Major}, {and} \bibinfo{person}{Shmargaret Shmitchell}.} \bibinfo{year}{2021}\natexlab{}.
\newblock \showarticletitle{On the dangers of stochastic parrots: Can language models be too big?}. In \bibinfo{booktitle}{\emph{Proceedings of the 2021 ACM conference on fairness, accountability, and transparency}}. \bibinfo{publisher}{{}}, \bibinfo{address}{{}}, \bibinfo{pages}{610--623}.
\newblock


\bibitem[Birhane et~al\mbox{.}(2022a)]%
        {birhane2022power}
\bibfield{author}{\bibinfo{person}{Abeba Birhane}, \bibinfo{person}{William Isaac}, \bibinfo{person}{Vinodkumar Prabhakaran}, \bibinfo{person}{Mark Diaz}, \bibinfo{person}{Madeleine~Clare Elish}, \bibinfo{person}{Iason Gabriel}, {and} \bibinfo{person}{Shakir Mohamed}.} \bibinfo{year}{2022}\natexlab{a}.
\newblock \showarticletitle{Power to the people? opportunities and challenges for participatory AI}.
\newblock \bibinfo{journal}{\emph{Equity and Access in Algorithms, Mechanisms, and Optimization}} (\bibinfo{year}{2022}), \bibinfo{pages}{1--8}.
\newblock


\bibitem[Birhane et~al\mbox{.}(2022b)]%
        {birhane2022values}
\bibfield{author}{\bibinfo{person}{Abeba Birhane}, \bibinfo{person}{Pratyusha Kalluri}, \bibinfo{person}{Dallas Card}, \bibinfo{person}{William Agnew}, \bibinfo{person}{Ravit Dotan}, {and} \bibinfo{person}{Michelle Bao}.} \bibinfo{year}{2022}\natexlab{b}.
\newblock \showarticletitle{The values encoded in machine learning research}. In \bibinfo{booktitle}{\emph{Proceedings of the 2022 ACM Conference on Fairness, Accountability, and Transparency}}. \bibinfo{publisher}{{}}, \bibinfo{address}{{}}, \bibinfo{pages}{173--184}.
\newblock


\bibitem[Bj\"{o}rgvinsson et~al\mbox{.}(2010)]%
        {10.1145/1900441.1900448}
\bibfield{author}{\bibinfo{person}{Erling Bj\"{o}rgvinsson}, \bibinfo{person}{Pelle Ehn}, {and} \bibinfo{person}{Per-Anders Hillgren}.} \bibinfo{year}{2010}\natexlab{}.
\newblock \showarticletitle{Participatory design and "democratizing innovation"}. In \bibinfo{booktitle}{\emph{Proceedings of the 11th Biennial Participatory Design Conference}} (Sydney, Australia) \emph{(\bibinfo{series}{PDC '10})}. \bibinfo{publisher}{Association for Computing Machinery}, \bibinfo{address}{New York, NY, USA}, \bibinfo{pages}{41–50}.
\newblock
\showISBNx{9781450301312}
\urldef\tempurl%
\url{https://doi.org/10.1145/1900441.1900448}
\showDOI{\tempurl}


\bibitem[Braun and Clarke(2022)]%
        {Braun2022EverythingCW}
\bibfield{author}{\bibinfo{person}{Virginia Braun} {and} \bibinfo{person}{Victoria Clarke}.} \bibinfo{year}{2022}\natexlab{}.
\newblock \showarticletitle{Everything changes… well some things do: Reflections on, and resources for, reflexive thematic analysis}.
\newblock \bibinfo{journal}{\emph{QMiP Bulletin}} (\bibinfo{year}{2022}).
\newblock
\urldef\tempurl%
\url{https://api.semanticscholar.org/CorpusID:255921405}
\showURL{%
\tempurl}


\bibitem[Brookes and Waller(2023)]%
        {brookes2023communities}
\bibfield{author}{\bibinfo{person}{Stephanie Brookes} {and} \bibinfo{person}{Lisa Waller}.} \bibinfo{year}{2023}\natexlab{}.
\newblock \showarticletitle{Communities of practice in the production and resourcing of fact-checking}.
\newblock \bibinfo{journal}{\emph{Journalism}} \bibinfo{volume}{24}, \bibinfo{number}{9} (\bibinfo{year}{2023}), \bibinfo{pages}{1938--1958}.
\newblock


\bibitem[Brown et~al\mbox{.}(2020)]%
        {brown2020language}
\bibfield{author}{\bibinfo{person}{Tom Brown}, \bibinfo{person}{Benjamin Mann}, \bibinfo{person}{Nick Ryder}, \bibinfo{person}{Melanie Subbiah}, \bibinfo{person}{Jared~D Kaplan}, \bibinfo{person}{Prafulla Dhariwal}, \bibinfo{person}{Arvind Neelakantan}, \bibinfo{person}{Pranav Shyam}, \bibinfo{person}{Girish Sastry}, \bibinfo{person}{Amanda Askell}, {et~al\mbox{.}}} \bibinfo{year}{2020}\natexlab{}.
\newblock \showarticletitle{Language models are few-shot learners}.
\newblock \bibinfo{journal}{\emph{Advances in neural information processing systems}}  \bibinfo{volume}{33} (\bibinfo{year}{2020}), \bibinfo{pages}{1877--1901}.
\newblock


\bibitem[Brynjolfsson et~al\mbox{.}(2021)]%
        {brynjolfsson2021productivity}
\bibfield{author}{\bibinfo{person}{Erik Brynjolfsson}, \bibinfo{person}{Daniel Rock}, {and} \bibinfo{person}{Chad Syverson}.} \bibinfo{year}{2021}\natexlab{}.
\newblock \showarticletitle{The productivity J-curve: How intangibles complement general purpose technologies}.
\newblock \bibinfo{journal}{\emph{American Economic Journal: Macroeconomics}} \bibinfo{volume}{13}, \bibinfo{number}{1} (\bibinfo{year}{2021}), \bibinfo{pages}{333--372}.
\newblock


\bibitem[Card et~al\mbox{.}(1990)]%
        {card1990design}
\bibfield{author}{\bibinfo{person}{Stuart~K Card}, \bibinfo{person}{Jock~D Mackinlay}, {and} \bibinfo{person}{George~G Robertson}.} \bibinfo{year}{1990}\natexlab{}.
\newblock \showarticletitle{The design space of input devices}. In \bibinfo{booktitle}{\emph{Proceedings of the SIGCHI conference on Human factors in computing systems}}. \bibinfo{pages}{117--124}.
\newblock


\bibitem[Center(2024)]%
        {tfc}
\bibfield{author}{\bibinfo{person}{Taiwan~FactCheck Center}.} \bibinfo{year}{2024}\natexlab{}.
\newblock \bibinfo{title}{Taiwan FactCheck Center}.
\newblock \bibinfo{howpublished}{\url{https://tfc-taiwan.org.tw/en}}.
\newblock
\newblock
\shownote{[Accessed 22-01-2024]}.


\bibitem[Chakrabarty et~al\mbox{.}(2023)]%
        {Chakrabarty2023CreativitySI}
\bibfield{author}{\bibinfo{person}{Tuhin Chakrabarty}, \bibinfo{person}{Vishakh Padmakumar}, \bibinfo{person}{Faeze Brahman}, {and} \bibinfo{person}{Smaranda Muresan}.} \bibinfo{year}{2023}\natexlab{}.
\newblock \showarticletitle{Creativity Support in the Age of Large Language Models: An Empirical Study Involving Emerging Writers}.
\newblock \bibinfo{journal}{\emph{ArXiv}}  \bibinfo{volume}{abs/2309.12570} (\bibinfo{year}{2023}).
\newblock
\urldef\tempurl%
\url{https://api.semanticscholar.org/CorpusID:262217523}
\showURL{%
\tempurl}


\bibitem[Check(2024)]%
        {africacheck}
\bibfield{author}{\bibinfo{person}{Africa Check}.} \bibinfo{year}{2024}\natexlab{}.
\newblock \bibinfo{title}{Africa Check}.
\newblock \bibinfo{howpublished}{\url{https://africacheck.org/}}.
\newblock
\newblock
\shownote{[Accessed 22-01-2024]}.


\bibitem[Chen et~al\mbox{.}(2021)]%
        {chen2021evaluating}
\bibfield{author}{\bibinfo{person}{Mark Chen}, \bibinfo{person}{Jerry Tworek}, \bibinfo{person}{Heewoo Jun}, \bibinfo{person}{Qiming Yuan}, \bibinfo{person}{Henrique Ponde de~Oliveira Pinto}, \bibinfo{person}{Jared Kaplan}, \bibinfo{person}{Harri Edwards}, \bibinfo{person}{Yuri Burda}, \bibinfo{person}{Nicholas Joseph}, \bibinfo{person}{Greg Brockman}, {et~al\mbox{.}}} \bibinfo{year}{2021}\natexlab{}.
\newblock \showarticletitle{Evaluating large language models trained on code}.
\newblock \bibinfo{journal}{\emph{arXiv preprint arXiv:2107.03374}} (\bibinfo{year}{2021}).
\newblock


\bibitem[Chequeado(2024)]%
        {chequeado}
\bibfield{author}{\bibinfo{person}{Chequeado}.} \bibinfo{year}{2024}\natexlab{}.
\newblock \bibinfo{title}{Chequeado}.
\newblock \bibinfo{howpublished}{\url{https://chequeado.com/}}.
\newblock
\newblock
\shownote{[Accessed 22-01-2024]}.


\bibitem[Clarke and Braun(2017)]%
        {Clarke2017ThematicA}
\bibfield{author}{\bibinfo{person}{Victoria Clarke} {and} \bibinfo{person}{Virginia Braun}.} \bibinfo{year}{2017}\natexlab{}.
\newblock \showarticletitle{Thematic analysis}.
\newblock \bibinfo{journal}{\emph{The Journal of Positive Psychology}}  \bibinfo{volume}{12} (\bibinfo{year}{2017}), \bibinfo{pages}{297 -- 298}.
\newblock
\urldef\tempurl%
\url{https://api.semanticscholar.org/CorpusID:219624951}
\showURL{%
\tempurl}


\bibitem[ColombiaCheck(2024)]%
        {colombiacheck}
\bibfield{author}{\bibinfo{person}{ColombiaCheck}.} \bibinfo{year}{2024}\natexlab{}.
\newblock \bibinfo{title}{ColombiaCheck}.
\newblock \bibinfo{howpublished}{\url{https://colombiacheck.com/}}.
\newblock
\newblock
\shownote{[Accessed 22-01-2024]}.


\bibitem[CrossCheck(2024)]%
        {rmit}
\bibfield{author}{\bibinfo{person}{RMIT~FactLab CrossCheck}.} \bibinfo{year}{2024}\natexlab{}.
\newblock \bibinfo{title}{RMIT FactLab CrossCheck}.
\newblock \bibinfo{howpublished}{\url{https://www.rmit.edu.au/about/schools-colleges/media-and-communication/industry/factlab/crosscheck}}.
\newblock
\newblock
\shownote{[Accessed 22-01-2024]}.


\bibitem[Das et~al\mbox{.}(2023)]%
        {das2023state}
\bibfield{author}{\bibinfo{person}{Anubrata Das}, \bibinfo{person}{Houjiang Liu}, \bibinfo{person}{Venelin Kovatchev}, {and} \bibinfo{person}{Matthew Lease}.} \bibinfo{year}{2023}\natexlab{}.
\newblock \showarticletitle{The state of human-centered NLP technology for fact-checking}.
\newblock \bibinfo{journal}{\emph{Information processing \& management}} \bibinfo{volume}{60}, \bibinfo{number}{2} (\bibinfo{year}{2023}), \bibinfo{pages}{103219}.
\newblock


\bibitem[Delgado et~al\mbox{.}(2023)]%
        {10.1145/3617694.3623261}
\bibfield{author}{\bibinfo{person}{Fernando Delgado}, \bibinfo{person}{Stephen Yang}, \bibinfo{person}{Michael Madaio}, {and} \bibinfo{person}{Qian Yang}.} \bibinfo{year}{2023}\natexlab{}.
\newblock \showarticletitle{The Participatory Turn in AI Design: Theoretical Foundations and the Current State of Practice}. In \bibinfo{booktitle}{\emph{Proceedings of the 3rd ACM Conference on Equity and Access in Algorithms, Mechanisms, and Optimization}} (<conf-loc>, <city>Boston</city>, <state>MA</state>, <country>USA</country>, </conf-loc>) \emph{(\bibinfo{series}{EAAMO '23})}. \bibinfo{publisher}{Association for Computing Machinery}, \bibinfo{address}{New York, NY, USA}, Article \bibinfo{articleno}{37}, \bibinfo{numpages}{23}~pages.
\newblock
\showISBNx{9798400703812}
\urldef\tempurl%
\url{https://doi.org/10.1145/3617694.3623261}
\showDOI{\tempurl}


\bibitem[Detector(2024)]%
        {eldetector}
\bibfield{author}{\bibinfo{person}{Univision~El Detector}.} \bibinfo{year}{2024}\natexlab{}.
\newblock \bibinfo{title}{Univision El Detector}.
\newblock \bibinfo{howpublished}{\url{https://www.univision.com/especiales/noticias/detector/index.html}}.
\newblock
\newblock
\shownote{[Accessed 22-01-2024]}.


\bibitem[Dierickx et~al\mbox{.}(2023)]%
        {dierickx2023automated}
\bibfield{author}{\bibinfo{person}{Laurence Dierickx}, \bibinfo{person}{Carl-Gustav Lind{\'e}n}, {and} \bibinfo{person}{Andreas~Lothe Opdahl}.} \bibinfo{year}{2023}\natexlab{}.
\newblock \showarticletitle{Automated fact-checking to support professional practices: systematic literature review and meta-analysis}.
\newblock \bibinfo{journal}{\emph{International Journal of Communication}}  \bibinfo{volume}{17} (\bibinfo{year}{2023}), \bibinfo{pages}{21}.
\newblock


\bibitem[Ekstr{\"o}m et~al\mbox{.}(2021)]%
        {ekstrom2021epistemologies}
\bibfield{author}{\bibinfo{person}{Mats Ekstr{\"o}m}, \bibinfo{person}{Amanda Rams{\"a}lv}, {and} \bibinfo{person}{Oscar Westlund}.} \bibinfo{year}{2021}\natexlab{}.
\newblock \showarticletitle{The epistemologies of breaking news}.
\newblock \bibinfo{journal}{\emph{Journalism Studies}} \bibinfo{volume}{22}, \bibinfo{number}{2} (\bibinfo{year}{2021}), \bibinfo{pages}{174--192}.
\newblock


\bibitem[Ekstr{\"o}m et~al\mbox{.}(2022)]%
        {ekstrom2022data}
\bibfield{author}{\bibinfo{person}{Mats Ekstr{\"o}m}, \bibinfo{person}{Amanda Rams{\"a}lv}, {and} \bibinfo{person}{Oscar Westlund}.} \bibinfo{year}{2022}\natexlab{}.
\newblock \showarticletitle{Data-driven news work culture: Reconciling tensions in epistemic values and practices of news journalism}.
\newblock \bibinfo{journal}{\emph{Journalism}} \bibinfo{volume}{23}, \bibinfo{number}{4} (\bibinfo{year}{2022}), \bibinfo{pages}{755--772}.
\newblock


\bibitem[Eloundou et~al\mbox{.}(2023)]%
        {eloundou2023gpts}
\bibfield{author}{\bibinfo{person}{Tyna Eloundou}, \bibinfo{person}{Sam Manning}, \bibinfo{person}{Pamela Mishkin}, {and} \bibinfo{person}{Daniel Rock}.} \bibinfo{year}{2023}\natexlab{}.
\newblock \showarticletitle{Gpts are gpts: An early look at the labor market impact potential of large language models}.
\newblock \bibinfo{journal}{\emph{arXiv preprint arXiv:2303.10130}} (\bibinfo{year}{2023}).
\newblock


\bibitem[Etikan et~al\mbox{.}(2016)]%
        {etikan2016comparison}
\bibfield{author}{\bibinfo{person}{Ilker Etikan}, \bibinfo{person}{Sulaiman~Abubakar Musa}, \bibinfo{person}{Rukayya~Sunusi Alkassim}, {et~al\mbox{.}}} \bibinfo{year}{2016}\natexlab{}.
\newblock \showarticletitle{Comparison of convenience sampling and purposive sampling}.
\newblock \bibinfo{journal}{\emph{American journal of theoretical and applied statistics}} \bibinfo{volume}{5}, \bibinfo{number}{1} (\bibinfo{year}{2016}), \bibinfo{pages}{1--4}.
\newblock


\bibitem[Factly(2024)]%
        {factly}
\bibfield{author}{\bibinfo{person}{Factly}.} \bibinfo{year}{2024}\natexlab{}.
\newblock \bibinfo{title}{Factly}.
\newblock \bibinfo{howpublished}{\url{https://factly.in/}}.
\newblock
\newblock
\shownote{[Accessed 22-01-2024]}.


\bibitem[Fatos(2024)]%
        {aosfatos}
\bibfield{author}{\bibinfo{person}{Aos Fatos}.} \bibinfo{year}{2024}\natexlab{}.
\newblock \bibinfo{title}{Aos Fatos}.
\newblock \bibinfo{howpublished}{\url{https://www.aosfatos.org/}}.
\newblock
\newblock
\shownote{[Accessed 22-01-2024]}.


\bibitem[Feedback(2024)]%
        {scifeedback}
\bibfield{author}{\bibinfo{person}{Science Feedback}.} \bibinfo{year}{2024}\natexlab{}.
\newblock \bibinfo{title}{Science Feedback}.
\newblock \bibinfo{howpublished}{\url{https://science.feedback.org/}}.
\newblock
\newblock
\shownote{[Accessed 22-01-2024]}.


\bibitem[Fichman and Kemerer(1999)]%
        {fichman1999illusory}
\bibfield{author}{\bibinfo{person}{Robert~G Fichman} {and} \bibinfo{person}{Chris~F Kemerer}.} \bibinfo{year}{1999}\natexlab{}.
\newblock \showarticletitle{The illusory diffusion of innovation: An examination of assimilation gaps}.
\newblock \bibinfo{journal}{\emph{Information systems research}} \bibinfo{volume}{10}, \bibinfo{number}{3} (\bibinfo{year}{1999}), \bibinfo{pages}{255--275}.
\newblock


\bibitem[for Africa(2024)]%
        {codeforafrica}
\bibfield{author}{\bibinfo{person}{Code for Africa}.} \bibinfo{year}{2024}\natexlab{}.
\newblock \bibinfo{title}{Code for Africa}.
\newblock \bibinfo{howpublished}{\url{https://github.com/CodeForAfrica/}}.
\newblock
\newblock
\shownote{[Accessed 22-01-2024]}.


\bibitem[Friedman et~al\mbox{.}(2017)]%
        {10.1561/1100000015}
\bibfield{author}{\bibinfo{person}{Batya Friedman}, \bibinfo{person}{David~G. Hendry}, {and} \bibinfo{person}{Alan Borning}.} \bibinfo{year}{2017}\natexlab{}.
\newblock \showarticletitle{A Survey of Value Sensitive Design Methods}.
\newblock \bibinfo{journal}{\emph{Found. Trends Hum.-Comput. Interact.}} \bibinfo{volume}{11}, \bibinfo{number}{2} (\bibinfo{date}{nov} \bibinfo{year}{2017}), \bibinfo{pages}{63–125}.
\newblock
\showISSN{1551-3955}
\urldef\tempurl%
\url{https://doi.org/10.1561/1100000015}
\showDOI{\tempurl}


\bibitem[Grammarly(2024)]%
        {grammarly}
\bibfield{author}{\bibinfo{person}{Grammarly}.} \bibinfo{year}{2024}\natexlab{}.
\newblock \bibinfo{title}{Grammarly}.
\newblock \bibinfo{howpublished}{\url{https://www.grammarly.com/}}.
\newblock
\newblock
\shownote{[Accessed 22-01-2024]}.


\bibitem[Graves(2017)]%
        {graves2017anatomy}
\bibfield{author}{\bibinfo{person}{Lucas Graves}.} \bibinfo{year}{2017}\natexlab{}.
\newblock \showarticletitle{Anatomy of a fact check: Objective practice and the contested epistemology of fact checking}.
\newblock \bibinfo{journal}{\emph{Communication, culture \& critique}} \bibinfo{volume}{10}, \bibinfo{number}{3} (\bibinfo{year}{2017}), \bibinfo{pages}{518--537}.
\newblock


\bibitem[Graves(2018)]%
        {graves2018factsheet}
\bibfield{author}{\bibinfo{person}{Lucas Graves}.} \bibinfo{year}{2018}\natexlab{}.
\newblock \showarticletitle{Factsheet: Understanding the promise and limits of automated fact-checking}.
\newblock \bibinfo{journal}{\emph{Reuters Inst. Study of Journalism, Univ. Oxford, Oxford}} (\bibinfo{year}{2018}).
\newblock


\bibitem[Graves and Amazeen(2019)]%
        {graves2019fact}
\bibfield{author}{\bibinfo{person}{Lucas Graves} {and} \bibinfo{person}{Michelle~A Amazeen}.} \bibinfo{year}{2019}\natexlab{}.
\newblock \showarticletitle{Fact-checking as idea and practice in journalism}.
\newblock In \bibinfo{booktitle}{\emph{Oxford research encyclopedia of communication}}.
\newblock


\bibitem[Guo et~al\mbox{.}(2022)]%
        {guo2022survey}
\bibfield{author}{\bibinfo{person}{Zhijiang Guo}, \bibinfo{person}{Michael Schlichtkrull}, {and} \bibinfo{person}{Andreas Vlachos}.} \bibinfo{year}{2022}\natexlab{}.
\newblock \showarticletitle{A survey on automated fact-checking}.
\newblock \bibinfo{journal}{\emph{Transactions of the Association for Computational Linguistics}}  \bibinfo{volume}{10} (\bibinfo{year}{2022}), \bibinfo{pages}{178--206}.
\newblock


\bibitem[Infoveritas(2024)]%
        {infoveritas}
\bibfield{author}{\bibinfo{person}{Infoveritas}.} \bibinfo{year}{2024}\natexlab{}.
\newblock \bibinfo{title}{Infoveritas}.
\newblock \bibinfo{howpublished}{\url{https://info-veritas.com/}}.
\newblock
\newblock
\shownote{[Accessed 22-01-2024]}.


\bibitem[Jain et~al\mbox{.}(2023)]%
        {jain2023ai}
\bibfield{author}{\bibinfo{person}{Shrey Jain}, \bibinfo{person}{Connor Spelliscy}, \bibinfo{person}{Samuel Vance-Law}, {and} \bibinfo{person}{Scott Moore}.} \bibinfo{year}{2023}\natexlab{}.
\newblock \showarticletitle{AI and Democracy's Digital Identity Crisis}.
\newblock \bibinfo{journal}{\emph{arXiv preprint arXiv:2311.16115}} (\bibinfo{year}{2023}).
\newblock


\bibitem[Jiang et~al\mbox{.}(2023)]%
        {jiang2023mistral}
\bibfield{author}{\bibinfo{person}{Albert~Q Jiang}, \bibinfo{person}{Alexandre Sablayrolles}, \bibinfo{person}{Arthur Mensch}, \bibinfo{person}{Chris Bamford}, \bibinfo{person}{Devendra~Singh Chaplot}, \bibinfo{person}{Diego de~las Casas}, \bibinfo{person}{Florian Bressand}, \bibinfo{person}{Gianna Lengyel}, \bibinfo{person}{Guillaume Lample}, \bibinfo{person}{Lucile Saulnier}, {et~al\mbox{.}}} \bibinfo{year}{2023}\natexlab{}.
\newblock \showarticletitle{Mistral 7B}.
\newblock \bibinfo{journal}{\emph{arXiv preprint arXiv:2310.06825}} \bibinfo{volume}{{}}, \bibinfo{number}{{}} (\bibinfo{year}{2023}), \bibinfo{pages}{{}}.
\newblock


\bibitem[Juneja and Mitra(2022)]%
        {juneja2022human}
\bibfield{author}{\bibinfo{person}{Prerna Juneja} {and} \bibinfo{person}{Tanushree Mitra}.} \bibinfo{year}{2022}\natexlab{}.
\newblock \showarticletitle{Human and technological infrastructures of fact-checking}.
\newblock \bibinfo{journal}{\emph{Proceedings of the ACM on Human-Computer Interaction}} \bibinfo{volume}{6}, \bibinfo{number}{CSCW2} (\bibinfo{year}{2022}), \bibinfo{pages}{1--36}.
\newblock


\bibitem[Kapoor and Narayanan(2023)]%
        {kapoor2023prepare}
\bibfield{author}{\bibinfo{person}{S Kapoor} {and} \bibinfo{person}{A Narayanan}.} \bibinfo{year}{2023}\natexlab{}.
\newblock \bibinfo{title}{How to prepare for the deluge of generative AI on social media}.
\newblock
\newblock


\bibitem[Li et~al\mbox{.}(2023)]%
        {10.1145/3593013.3594070}
\bibfield{author}{\bibinfo{person}{Hanlin Li}, \bibinfo{person}{Nicholas Vincent}, \bibinfo{person}{Stevie Chancellor}, {and} \bibinfo{person}{Brent Hecht}.} \bibinfo{year}{2023}\natexlab{}.
\newblock \showarticletitle{The Dimensions of Data Labor: A Road Map for Researchers, Activists, and Policymakers to Empower Data Producers}. In \bibinfo{booktitle}{\emph{Proceedings of the 2023 ACM Conference on Fairness, Accountability, and Transparency}} (Chicago, IL, USA) \emph{(\bibinfo{series}{FAccT '23})}. \bibinfo{publisher}{Association for Computing Machinery}, \bibinfo{address}{New York, NY, USA}, \bibinfo{pages}{1151–1161}.
\newblock
\showISBNx{9798400701924}
\urldef\tempurl%
\url{https://doi.org/10.1145/3593013.3594070}
\showDOI{\tempurl}


\bibitem[Litmus(2024)]%
        {litmus}
\bibfield{author}{\bibinfo{person}{Litmus}.} \bibinfo{year}{2024}\natexlab{}.
\newblock \bibinfo{title}{Litmus}.
\newblock \bibinfo{howpublished}{\url{https://litmus-factcheck.jp/about/en/}}.
\newblock
\newblock
\shownote{[Accessed 22-01-2024]}.


\bibitem[logically.ai(2024)]%
        {logically}
\bibfield{author}{\bibinfo{person}{logically.ai}.} \bibinfo{year}{2024}\natexlab{}.
\newblock \bibinfo{title}{logically.ai}.
\newblock \bibinfo{howpublished}{\url{https://www.logically.ai/}}.
\newblock
\newblock
\shownote{[Accessed 22-01-2024]}.


\bibitem[Longoni et~al\mbox{.}(2022)]%
        {longoni2022news}
\bibfield{author}{\bibinfo{person}{Chiara Longoni}, \bibinfo{person}{Andrey Fradkin}, \bibinfo{person}{Luca Cian}, {and} \bibinfo{person}{Gordon Pennycook}.} \bibinfo{year}{2022}\natexlab{}.
\newblock \showarticletitle{News from generative artificial intelligence is believed less}. In \bibinfo{booktitle}{\emph{Proceedings of the 2022 ACM Conference on Fairness, Accountability, and Transparency}}. \bibinfo{pages}{97--106}.
\newblock


\bibitem[Maldita(2024)]%
        {maldita}
\bibfield{author}{\bibinfo{person}{Maldita}.} \bibinfo{year}{2024}\natexlab{}.
\newblock \bibinfo{title}{Maldita}.
\newblock \bibinfo{howpublished}{\url{https://maldita.es/}}.
\newblock
\newblock
\shownote{[Accessed 22-01-2024]}.


\bibitem[Meedan(2024)]%
        {meedan}
\bibfield{author}{\bibinfo{person}{Meedan}.} \bibinfo{year}{2024}\natexlab{}.
\newblock \bibinfo{title}{Meedan}.
\newblock \bibinfo{howpublished}{\url{https://meedan.com/}}.
\newblock
\newblock
\shownote{[Accessed 22-01-2024]}.


\bibitem[MindaNews(2024)]%
        {mindanews}
\bibfield{author}{\bibinfo{person}{MindaNews}.} \bibinfo{year}{2024}\natexlab{}.
\newblock \bibinfo{title}{MindaNews}.
\newblock \bibinfo{howpublished}{\url{https://www.mindanews.com/}}.
\newblock
\newblock
\shownote{[Accessed 22-01-2024]}.


\bibitem[Morris(2023)]%
        {morris2023scientists}
\bibfield{author}{\bibinfo{person}{Meredith~Ringel Morris}.} \bibinfo{year}{2023}\natexlab{}.
\newblock \showarticletitle{Scientists' Perspectives on the Potential for Generative AI in their Fields}.
\newblock \bibinfo{journal}{\emph{arXiv preprint arXiv:2304.01420}} (\bibinfo{year}{2023}).
\newblock


\bibitem[Morris et~al\mbox{.}(2023)]%
        {morris2023design}
\bibfield{author}{\bibinfo{person}{Meredith~Ringel Morris}, \bibinfo{person}{Carrie~J Cai}, \bibinfo{person}{Jess Holbrook}, \bibinfo{person}{Chinmay Kulkarni}, {and} \bibinfo{person}{Michael Terry}.} \bibinfo{year}{2023}\natexlab{}.
\newblock \showarticletitle{The design space of generative models}.
\newblock \bibinfo{journal}{\emph{arXiv preprint arXiv:2304.10547}} (\bibinfo{year}{2023}).
\newblock


\bibitem[Naderifar et~al\mbox{.}(2017)]%
        {naderifar2017snowball}
\bibfield{author}{\bibinfo{person}{Mahin Naderifar}, \bibinfo{person}{Hamideh Goli}, {and} \bibinfo{person}{Fereshteh Ghaljaie}.} \bibinfo{year}{2017}\natexlab{}.
\newblock \showarticletitle{Snowball sampling: A purposeful method of sampling in qualitative research}.
\newblock \bibinfo{journal}{\emph{Strides in development of medical education}} \bibinfo{volume}{14}, \bibinfo{number}{3} (\bibinfo{year}{2017}).
\newblock


\bibitem[Neumann and Wolczynski(2023)]%
        {neumann2023does}
\bibfield{author}{\bibinfo{person}{Terrence Neumann} {and} \bibinfo{person}{Nicholas Wolczynski}.} \bibinfo{year}{2023}\natexlab{}.
\newblock \showarticletitle{Does AI-Assisted Fact-Checking Disproportionately Benefit Majority Groups Online?}. In \bibinfo{booktitle}{\emph{Proceedings of the 2023 ACM Conference on Fairness, Accountability, and Transparency}}. \bibinfo{pages}{480--490}.
\newblock


\bibitem[Newtral(2024)]%
        {newtral}
\bibfield{author}{\bibinfo{person}{Newtral}.} \bibinfo{year}{2024}\natexlab{}.
\newblock \bibinfo{title}{Newtral}.
\newblock \bibinfo{howpublished}{\url{https://www.newtral.es/}}.
\newblock
\newblock
\shownote{[Accessed 22-01-2024]}.


\bibitem[OpenAI(2022)]%
        {openai2022chatgpt}
\bibfield{author}{\bibinfo{person}{OpenAI}.} \bibinfo{year}{2022}\natexlab{}.
\newblock \showarticletitle{Introducing ChatGPT}.
\newblock \bibinfo{journal}{\emph{OpenAI Blog}} \bibinfo{volume}{{}}, \bibinfo{number}{{}} (\bibinfo{date}{Nov} \bibinfo{year}{2022}), \bibinfo{pages}{{}}.
\newblock


\bibitem[Ouyang et~al\mbox{.}(2022)]%
        {ouyang2022training}
\bibfield{author}{\bibinfo{person}{Long Ouyang}, \bibinfo{person}{Jeffrey Wu}, \bibinfo{person}{Xu Jiang}, \bibinfo{person}{Diogo Almeida}, \bibinfo{person}{Carroll Wainwright}, \bibinfo{person}{Pamela Mishkin}, \bibinfo{person}{Chong Zhang}, \bibinfo{person}{Sandhini Agarwal}, \bibinfo{person}{Katarina Slama}, \bibinfo{person}{Alex Ray}, {et~al\mbox{.}}} \bibinfo{year}{2022}\natexlab{}.
\newblock \showarticletitle{Training language models to follow instructions with human feedback}.
\newblock \bibinfo{journal}{\emph{Advances in Neural Information Processing Systems}}  \bibinfo{volume}{35} (\bibinfo{year}{2022}), \bibinfo{pages}{27730--27744}.
\newblock


\bibitem[Phillips(2010)]%
        {doi:10.1080/17512781003642972}
\bibfield{author}{\bibinfo{person}{Angela Phillips}.} \bibinfo{year}{2010}\natexlab{}.
\newblock \showarticletitle{TRANSPARENCY AND THE NEW ETHICS OF JOURNALISM}.
\newblock \bibinfo{journal}{\emph{Journalism Practice}} \bibinfo{volume}{4}, \bibinfo{number}{3} (\bibinfo{year}{2010}), \bibinfo{pages}{373--382}.
\newblock
\urldef\tempurl%
\url{https://doi.org/10.1080/17512781003642972}
\showDOI{\tempurl}
\showeprint{https://doi.org/10.1080/17512781003642972}


\bibitem[Politica(2024)]%
        {pagella}
\bibfield{author}{\bibinfo{person}{Pagella Politica}.} \bibinfo{year}{2024}\natexlab{}.
\newblock \bibinfo{title}{Pagella Politica}.
\newblock \bibinfo{howpublished}{\url{https://pagellapolitica.it/}}.
\newblock
\newblock
\shownote{[Accessed 22-01-2024]}.


\bibitem[Politifact(2024)]%
        {politifact}
\bibfield{author}{\bibinfo{person}{Politifact}.} \bibinfo{year}{2024}\natexlab{}.
\newblock \bibinfo{title}{Politifact}.
\newblock \bibinfo{howpublished}{\url{https://www.politifact.com/}}.
\newblock
\newblock
\shownote{[Accessed 22-01-2024]}.


\bibitem[Poynter(2024a)]%
        {ifcn}
\bibfield{author}{\bibinfo{person}{Poynter}.} \bibinfo{year}{2024}\natexlab{a}.
\newblock \bibinfo{title}{International Fact Checking Network}.
\newblock \bibinfo{howpublished}{\url{https://www.poynter.org/ifcn/}}.
\newblock
\newblock
\shownote{[Accessed 22-01-2024]}.


\bibitem[Poynter(2024b)]%
        {ifcnsignatories}
\bibfield{author}{\bibinfo{person}{Poynter}.} \bibinfo{year}{2024}\natexlab{b}.
\newblock \bibinfo{title}{Verified signatories of the IFCN code of principles}.
\newblock \bibinfo{howpublished}{\url{https://ifcncodeofprinciples.poynter.org/signatories}}.
\newblock
\newblock
\shownote{[Accessed 22-01-2024]}.


\bibitem[Prasad~Agrawal(2023)]%
        {prasad2023towards}
\bibfield{author}{\bibinfo{person}{Kalyan Prasad~Agrawal}.} \bibinfo{year}{2023}\natexlab{}.
\newblock \showarticletitle{Towards adoption of generative AI in organizational settings}.
\newblock \bibinfo{journal}{\emph{Journal of Computer Information Systems}} (\bibinfo{year}{2023}), \bibinfo{pages}{1--16}.
\newblock


\bibitem[Pravda(2024)]%
        {pravda}
\bibfield{author}{\bibinfo{person}{Pravda}.} \bibinfo{year}{2024}\natexlab{}.
\newblock \bibinfo{title}{Pravda}.
\newblock \bibinfo{howpublished}{\url{https://pravda.org.pl/}}.
\newblock
\newblock
\shownote{[Accessed 22-01-2024]}.


\bibitem[Press(2024)]%
        {aap}
\bibfield{author}{\bibinfo{person}{Australian~Associated Press}.} \bibinfo{year}{2024}\natexlab{}.
\newblock \bibinfo{title}{Australian Associated Press}.
\newblock \bibinfo{howpublished}{\url{https://www.aap.com.au/}}.
\newblock
\newblock
\shownote{[Accessed 22-01-2024]}.


\bibitem[Presse(2024)]%
        {afp}
\bibfield{author}{\bibinfo{person}{Agence~France Presse}.} \bibinfo{year}{2024}\natexlab{}.
\newblock \bibinfo{title}{Agence France Presse}.
\newblock \bibinfo{howpublished}{\url{https://www.afp.com/en}}.
\newblock
\newblock
\shownote{[Accessed 22-01-2024]}.


\bibitem[Queerinai et~al\mbox{.}(2023)]%
        {10.1145/3593013.3594134}
\bibfield{author}{\bibinfo{person}{Organizers~Of Queerinai}, \bibinfo{person}{Anaelia Ovalle}, \bibinfo{person}{Arjun Subramonian}, \bibinfo{person}{Ashwin Singh}, \bibinfo{person}{Claas Voelcker}, \bibinfo{person}{Danica~J. Sutherland}, \bibinfo{person}{Davide Locatelli}, \bibinfo{person}{Eva Breznik}, \bibinfo{person}{Filip Klubicka}, \bibinfo{person}{Hang Yuan}, \bibinfo{person}{Hetvi J}, \bibinfo{person}{Huan Zhang}, \bibinfo{person}{Jaidev Shriram}, \bibinfo{person}{Kruno Lehman}, \bibinfo{person}{Luca Soldaini}, \bibinfo{person}{Maarten Sap}, \bibinfo{person}{Marc~Peter Deisenroth}, \bibinfo{person}{Maria~Leonor Pacheco}, \bibinfo{person}{Maria Ryskina}, \bibinfo{person}{Martin Mundt}, \bibinfo{person}{Milind Agarwal}, \bibinfo{person}{Nyx Mclean}, \bibinfo{person}{Pan Xu}, \bibinfo{person}{A Pranav}, \bibinfo{person}{Raj Korpan}, \bibinfo{person}{Ruchira Ray}, \bibinfo{person}{Sarah Mathew}, \bibinfo{person}{Sarthak Arora}, \bibinfo{person}{St John}, \bibinfo{person}{Tanvi Anand},
  \bibinfo{person}{Vishakha Agrawal}, \bibinfo{person}{William Agnew}, \bibinfo{person}{Yanan Long}, \bibinfo{person}{Zijie~J. Wang}, \bibinfo{person}{Zeerak Talat}, \bibinfo{person}{Avijit Ghosh}, \bibinfo{person}{Nathaniel Dennler}, \bibinfo{person}{Michael Noseworthy}, \bibinfo{person}{Sharvani Jha}, \bibinfo{person}{Emi Baylor}, \bibinfo{person}{Aditya Joshi}, \bibinfo{person}{Natalia~Y. Bilenko}, \bibinfo{person}{Andrew Mcnamara}, \bibinfo{person}{Raphael Gontijo-Lopes}, \bibinfo{person}{Alex Markham}, \bibinfo{person}{Evyn Dong}, \bibinfo{person}{Jackie Kay}, \bibinfo{person}{Manu Saraswat}, \bibinfo{person}{Nikhil Vytla}, {and} \bibinfo{person}{Luke Stark}.} \bibinfo{year}{2023}\natexlab{}.
\newblock \showarticletitle{Queer In AI: A Case Study in Community-Led Participatory AI}. In \bibinfo{booktitle}{\emph{Proceedings of the 2023 ACM Conference on Fairness, Accountability, and Transparency}} (Chicago, IL, USA) \emph{(\bibinfo{series}{FAccT '23})}. \bibinfo{publisher}{Association for Computing Machinery}, \bibinfo{address}{New York, NY, USA}, \bibinfo{pages}{1882–1895}.
\newblock
\showISBNx{9798400701924}
\urldef\tempurl%
\url{https://doi.org/10.1145/3593013.3594134}
\showDOI{\tempurl}


\bibitem[Quint(2024)]%
        {quint}
\bibfield{author}{\bibinfo{person}{The Quint}.} \bibinfo{year}{2024}\natexlab{}.
\newblock \bibinfo{title}{The Quint}.
\newblock \bibinfo{howpublished}{\url{https://www.thequint.com/international}}.
\newblock
\newblock
\shownote{[Accessed 22-01-2024]}.


\bibitem[Radford et~al\mbox{.}(2018)]%
        {radford2018improving}
\bibfield{author}{\bibinfo{person}{Alec Radford}, \bibinfo{person}{Karthik Narasimhan}, \bibinfo{person}{Tim Salimans}, \bibinfo{person}{Ilya Sutskever}, {et~al\mbox{.}}} \bibinfo{year}{2018}\natexlab{}.
\newblock \showarticletitle{Improving language understanding by generative pre-training}.
\newblock \bibinfo{journal}{\emph{{}}} \bibinfo{volume}{{}}, \bibinfo{number}{{}} (\bibinfo{year}{2018}), \bibinfo{pages}{{}}.
\newblock


\bibitem[Radford et~al\mbox{.}(2019)]%
        {radford2019language}
\bibfield{author}{\bibinfo{person}{Alec Radford}, \bibinfo{person}{Jeffrey Wu}, \bibinfo{person}{Rewon Child}, \bibinfo{person}{David Luan}, \bibinfo{person}{Dario Amodei}, \bibinfo{person}{Ilya Sutskever}, {et~al\mbox{.}}} \bibinfo{year}{2019}\natexlab{}.
\newblock \showarticletitle{Language models are unsupervised multitask learners}.
\newblock \bibinfo{journal}{\emph{OpenAI blog}} \bibinfo{volume}{1}, \bibinfo{number}{8} (\bibinfo{year}{2019}), \bibinfo{pages}{9}.
\newblock


\bibitem[Radiya-Dixit and Neff(2023)]%
        {radiya2023sociotechnical}
\bibfield{author}{\bibinfo{person}{Evani Radiya-Dixit} {and} \bibinfo{person}{Gina Neff}.} \bibinfo{year}{2023}\natexlab{}.
\newblock \showarticletitle{A Sociotechnical Audit: Assessing Police Use of Facial Recognition}. In \bibinfo{booktitle}{\emph{Proceedings of the 2023 ACM Conference on Fairness, Accountability, and Transparency}}. \bibinfo{pages}{1334--1346}.
\newblock


\bibitem[Ramesh et~al\mbox{.}(2022)]%
        {ramesh2022hierarchical}
\bibfield{author}{\bibinfo{person}{Aditya Ramesh}, \bibinfo{person}{Prafulla Dhariwal}, \bibinfo{person}{Alex Nichol}, \bibinfo{person}{Casey Chu}, {and} \bibinfo{person}{Mark Chen}.} \bibinfo{year}{2022}\natexlab{}.
\newblock \showarticletitle{Hierarchical text-conditional image generation with clip latents}.
\newblock \bibinfo{journal}{\emph{arXiv preprint arXiv:2204.06125}} \bibinfo{volume}{1}, \bibinfo{number}{2} (\bibinfo{year}{2022}), \bibinfo{pages}{3}.
\newblock


\bibitem[Ramesh et~al\mbox{.}(2021)]%
        {ramesh2021zero}
\bibfield{author}{\bibinfo{person}{Aditya Ramesh}, \bibinfo{person}{Mikhail Pavlov}, \bibinfo{person}{Gabriel Goh}, \bibinfo{person}{Scott Gray}, \bibinfo{person}{Chelsea Voss}, \bibinfo{person}{Alec Radford}, \bibinfo{person}{Mark Chen}, {and} \bibinfo{person}{Ilya Sutskever}.} \bibinfo{year}{2021}\natexlab{}.
\newblock \showarticletitle{Zero-shot text-to-image generation}. In \bibinfo{booktitle}{\emph{International Conference on Machine Learning}}. PMLR, \bibinfo{pages}{8821--8831}.
\newblock


\bibitem[Rappler(2024)]%
        {rappler}
\bibfield{author}{\bibinfo{person}{Rappler}.} \bibinfo{year}{2024}\natexlab{}.
\newblock \bibinfo{title}{Rappler}.
\newblock \bibinfo{howpublished}{\url{https://www.rappler.com/}}.
\newblock
\newblock
\shownote{[Accessed 22-01-2024]}.


\bibitem[Reuters(2024)]%
        {reuters}
\bibfield{author}{\bibinfo{person}{Thomson Reuters}.} \bibinfo{year}{2024}\natexlab{}.
\newblock \bibinfo{title}{Thomson Reuters}.
\newblock \bibinfo{howpublished}{\url{https://www.thomsonreuters.com/en.html}}.
\newblock
\newblock
\shownote{[Accessed 22-01-2024]}.


\bibitem[Ritala et~al\mbox{.}(2023)]%
        {ritala2023transforming}
\bibfield{author}{\bibinfo{person}{Paavo Ritala}, \bibinfo{person}{Mika Ruokonen}, {and} \bibinfo{person}{Laavanya Ramaul}.} \bibinfo{year}{2023}\natexlab{}.
\newblock \showarticletitle{Transforming boundaries: how does ChatGPT change knowledge work?}
\newblock \bibinfo{journal}{\emph{Journal of Business Strategy}} \bibinfo{number}{ahead-of-print} (\bibinfo{year}{2023}).
\newblock


\bibitem[Rombach et~al\mbox{.}(2022)]%
        {rombach2022high}
\bibfield{author}{\bibinfo{person}{Robin Rombach}, \bibinfo{person}{Andreas Blattmann}, \bibinfo{person}{Dominik Lorenz}, \bibinfo{person}{Patrick Esser}, {and} \bibinfo{person}{Bj{\"o}rn Ommer}.} \bibinfo{year}{2022}\natexlab{}.
\newblock \showarticletitle{High-resolution image synthesis with latent diffusion models}. In \bibinfo{booktitle}{\emph{Proceedings of the IEEE/CVF conference on computer vision and pattern recognition}}. \bibinfo{pages}{10684--10695}.
\newblock


\bibitem[Singer et~al\mbox{.}(2022)]%
        {singer2022make}
\bibfield{author}{\bibinfo{person}{Uriel Singer}, \bibinfo{person}{Adam Polyak}, \bibinfo{person}{Thomas Hayes}, \bibinfo{person}{Xi Yin}, \bibinfo{person}{Jie An}, \bibinfo{person}{Songyang Zhang}, \bibinfo{person}{Qiyuan Hu}, \bibinfo{person}{Harry Yang}, \bibinfo{person}{Oron Ashual}, \bibinfo{person}{Oran Gafni}, {et~al\mbox{.}}} \bibinfo{year}{2022}\natexlab{}.
\newblock \showarticletitle{Make-a-video: Text-to-video generation without text-video data}.
\newblock \bibinfo{journal}{\emph{arXiv preprint arXiv:2209.14792}} (\bibinfo{year}{2022}).
\newblock


\bibitem[Spiegel(2024)]%
        {derspiegel}
\bibfield{author}{\bibinfo{person}{Der Spiegel}.} \bibinfo{year}{2024}\natexlab{}.
\newblock \bibinfo{title}{Der Spiegel}.
\newblock \bibinfo{howpublished}{\url{https://www.spiegel.de/international/}}.
\newblock
\newblock
\shownote{[Accessed 22-01-2024]}.


\bibitem[Spinuzzi(2005)]%
        {participatorydesign}
\bibfield{author}{\bibinfo{person}{Clay Spinuzzi}.} \bibinfo{year}{2005}\natexlab{}.
\newblock \showarticletitle{The Methodology of Participatory Design}.
\newblock \bibinfo{journal}{\emph{Technical Communication}}  \bibinfo{volume}{52} (\bibinfo{date}{05} \bibinfo{year}{2005}), \bibinfo{pages}{163--174}.
\newblock


\bibitem[Stories(2024)]%
        {leadstories}
\bibfield{author}{\bibinfo{person}{Lead Stories}.} \bibinfo{year}{2024}\natexlab{}.
\newblock \bibinfo{title}{Lead Stories}.
\newblock \bibinfo{howpublished}{\url{https://leadstories.com/}}.
\newblock
\newblock
\shownote{[Accessed 22-01-2024]}.


\bibitem[Tang et~al\mbox{.}(2015)]%
        {tang2015restructuring}
\bibfield{author}{\bibinfo{person}{Charlotte Tang}, \bibinfo{person}{Yunan Chen}, \bibinfo{person}{Bryan~C Semaan}, {and} \bibinfo{person}{Jahmeilah~A Roberson}.} \bibinfo{year}{2015}\natexlab{}.
\newblock \showarticletitle{Restructuring human infrastructure: The impact of EHR deployment in a volunteer-dependent clinic}. In \bibinfo{booktitle}{\emph{Proceedings of the 18th ACM Conference on Computer Supported Cooperative Work \& Social Computing}}. \bibinfo{pages}{649--661}.
\newblock


\bibitem[Tech4Peace(2024)]%
        {t4p}
\bibfield{author}{\bibinfo{person}{Tech4Peace}.} \bibinfo{year}{2024}\natexlab{}.
\newblock \bibinfo{title}{Tech4Peace}.
\newblock \bibinfo{howpublished}{\url{https://t4p.co/}}.
\newblock
\newblock
\shownote{[Accessed 22-01-2024]}.


\bibitem[Times(2023)]%
        {chatlaw}
\bibfield{author}{\bibinfo{person}{The New~York Times}.} \bibinfo{year}{2023}\natexlab{}.
\newblock \bibinfo{title}{The ChatGPT Lawyer Explains Himself}.
\newblock \bibinfo{howpublished}{\url{https://www.nytimes.com/2023/06/08/nyregion/lawyer-chatgpt-sanctions.html}}.
\newblock
\newblock
\shownote{[Accessed 22-01-2024]}.


\bibitem[Today(2024)]%
        {indiatoday}
\bibfield{author}{\bibinfo{person}{India Today}.} \bibinfo{year}{2024}\natexlab{}.
\newblock \bibinfo{title}{India Today}.
\newblock \bibinfo{howpublished}{\url{https://www.indiatoday.in/}}.
\newblock
\newblock
\shownote{[Accessed 22-01-2024]}.


\bibitem[Touvron et~al\mbox{.}(2023a)]%
        {touvron2023llama}
\bibfield{author}{\bibinfo{person}{Hugo Touvron}, \bibinfo{person}{Thibaut Lavril}, \bibinfo{person}{Gautier Izacard}, \bibinfo{person}{Xavier Martinet}, \bibinfo{person}{Marie-Anne Lachaux}, \bibinfo{person}{Timoth{\'e}e Lacroix}, \bibinfo{person}{Baptiste Rozi{\`e}re}, \bibinfo{person}{Naman Goyal}, \bibinfo{person}{Eric Hambro}, \bibinfo{person}{Faisal Azhar}, {et~al\mbox{.}}} \bibinfo{year}{2023}\natexlab{a}.
\newblock \showarticletitle{Llama: Open and efficient foundation language models}.
\newblock \bibinfo{journal}{\emph{arXiv preprint arXiv:2302.13971}} \bibinfo{volume}{{}}, \bibinfo{number}{{}} (\bibinfo{year}{2023}), \bibinfo{pages}{{}}.
\newblock


\bibitem[Touvron et~al\mbox{.}(2023b)]%
        {touvron2023llama2}
\bibfield{author}{\bibinfo{person}{Hugo Touvron}, \bibinfo{person}{Louis Martin}, \bibinfo{person}{Kevin Stone}, \bibinfo{person}{Peter Albert}, \bibinfo{person}{Amjad Almahairi}, \bibinfo{person}{Yasmine Babaei}, \bibinfo{person}{Nikolay Bashlykov}, \bibinfo{person}{Soumya Batra}, \bibinfo{person}{Prajjwal Bhargava}, \bibinfo{person}{Shruti Bhosale}, {et~al\mbox{.}}} \bibinfo{year}{2023}\natexlab{b}.
\newblock \showarticletitle{Llama 2: Open foundation and fine-tuned chat models}.
\newblock \bibinfo{journal}{\emph{arXiv preprint arXiv:2307.09288}} \bibinfo{volume}{{}}, \bibinfo{number}{{}} (\bibinfo{year}{2023}), \bibinfo{pages}{{}}.
\newblock


\bibitem[Wolk et~al\mbox{.}(2007)]%
        {wolk2007}
\bibfield{author}{\bibinfo{person}{Jessica Wolk}, \bibinfo{person}{Batya Friedman}, {and} \bibinfo{person}{Gavin Jancke}.} \bibinfo{year}{2007}\natexlab{}.
\newblock \showarticletitle{Value Tensions in Design: The Value Sensitive Design, Development, and Appropriation of a Corporation’s}.
\newblock \bibinfo{journal}{\emph{GROUP'07 - Proceedings of the 2007 International ACM Conference on Supporting Group Work}}, \bibinfo{pages}{281--290}.
\newblock
\urldef\tempurl%
\url{https://doi.org/10.1145/1316624.1316668}
\showDOI{\tempurl}


\bibitem[Wu et~al\mbox{.}(2023)]%
        {wu2023tune}
\bibfield{author}{\bibinfo{person}{Jay~Zhangjie Wu}, \bibinfo{person}{Yixiao Ge}, \bibinfo{person}{Xintao Wang}, \bibinfo{person}{Stan~Weixian Lei}, \bibinfo{person}{Yuchao Gu}, \bibinfo{person}{Yufei Shi}, \bibinfo{person}{Wynne Hsu}, \bibinfo{person}{Ying Shan}, \bibinfo{person}{Xiaohu Qie}, {and} \bibinfo{person}{Mike~Zheng Shou}.} \bibinfo{year}{2023}\natexlab{}.
\newblock \showarticletitle{Tune-a-video: One-shot tuning of image diffusion models for text-to-video generation}. In \bibinfo{booktitle}{\emph{Proceedings of the IEEE/CVF International Conference on Computer Vision}}. \bibinfo{pages}{7623--7633}.
\newblock


\bibitem[Xu et~al\mbox{.}(2022)]%
        {xu2022systematic}
\bibfield{author}{\bibinfo{person}{Frank~F Xu}, \bibinfo{person}{Uri Alon}, \bibinfo{person}{Graham Neubig}, {and} \bibinfo{person}{Vincent~Josua Hellendoorn}.} \bibinfo{year}{2022}\natexlab{}.
\newblock \showarticletitle{A systematic evaluation of large language models of code}. In \bibinfo{booktitle}{\emph{Proceedings of the 6th ACM SIGPLAN International Symposium on Machine Programming}}. \bibinfo{pages}{1--10}.
\newblock


\bibitem[Yu et~al\mbox{.}(2023)]%
        {yu2023antecedents}
\bibfield{author}{\bibinfo{person}{Xinying Yu}, \bibinfo{person}{Shi Xu}, {and} \bibinfo{person}{Mark Ashton}.} \bibinfo{year}{2023}\natexlab{}.
\newblock \showarticletitle{Antecedents and outcomes of artificial intelligence adoption and application in the workplace: the socio-technical system theory perspective}.
\newblock \bibinfo{journal}{\emph{Information Technology \& People}} \bibinfo{volume}{36}, \bibinfo{number}{1} (\bibinfo{year}{2023}), \bibinfo{pages}{454--474}.
\newblock


\bibitem[Zagni and Canetta(2023)]%
        {zagni2023generative}
\bibfield{author}{\bibinfo{person}{G Zagni} {and} \bibinfo{person}{T Canetta}.} \bibinfo{year}{2023}\natexlab{}.
\newblock \bibinfo{title}{Generative AI marks the beginning of a new era for disinformation}.
\newblock
\newblock


\bibitem[Zaj{\k{a}}c et~al\mbox{.}(2023)]%
        {zajkac2023clinician}
\bibfield{author}{\bibinfo{person}{Hubert~D Zaj{\k{a}}c}, \bibinfo{person}{Dana Li}, \bibinfo{person}{Xiang Dai}, \bibinfo{person}{Jonathan~F Carlsen}, \bibinfo{person}{Finn Kensing}, {and} \bibinfo{person}{Tariq~O Andersen}.} \bibinfo{year}{2023}\natexlab{}.
\newblock \showarticletitle{Clinician-facing AI in the Wild: Taking Stock of the Sociotechnical Challenges and Opportunities for HCI}.
\newblock \bibinfo{journal}{\emph{ACM Transactions on Computer-Human Interaction}} \bibinfo{volume}{30}, \bibinfo{number}{2} (\bibinfo{year}{2023}), \bibinfo{pages}{1--39}.
\newblock


\end{thebibliography}

\appendix

\end{document}